\documentclass[a4paper]{article}
\usepackage{amsmath,amsfonts,epsfig,color}

\numberwithin{equation}{section}
\newcommand{\hs}[1]{\hspace*{#1cm}}

\newcommand{\barr}{\begin{array}}
\newcommand{\earr}{\end{array}}
\newcommand{\bea}{\begin{eqnarray}}
\newcommand{\eea}{\end{eqnarray}}
\newcommand{\beqa}{\be \begin{array}{rcl}}
\newcommand{\eeqa}{\end{array} \ee}
\newcommand{\nn}{\nonumber}
\newcommand{\eqv}{=}

\newcommand{\et}[1]{e^{\mbox{\small $#1$}}}

\newcommand{\dt}{\! \cdot \!}
\newcommand{\wdg}{\! \wedge \!}
\newcommand{\crs}{\! \times \!}

\newcommand{\la}{\langle}
\newcommand{\ra}{\rangle}

\newcommand{\half}{{\textstyle \frac{1}{2}}}

\newcommand{\qrt}{{\textstyle \frac{1}{4}}}

\newcommand{\etal}{\textit{et al.}}

\newcommand{\alp}{\alpha}
\newcommand{\bet}{\beta}
\newcommand{\gam}{\gamma}
\newcommand{\del}{\delta}
\newcommand{\eps}{\epsilon}

\newcommand{\lam}{\lambda}
\newcommand{\sig}{\sigma}
\newcommand{\om}{\omega}

\newcommand{\Om}{\Omega}

\newcommand{\ba}{\mbox{\boldmath $a$}}
\newcommand{\bb}{\mbox{\boldmath $b$}}

\newcommand{\blde}{\mbox{\boldmath $e$}}

\newcommand{\bn}{\mbox{\boldmath $n$}}

\newcommand{\bx}{\mbox{\boldmath $x$}}

\newcommand{\bE}{\mbox{\boldmath $E$}}

\newcommand{\bsig}{\mbox{\boldmath $\sig$}}

\newcommand{\ldot}{\dot{l}}
\newcommand{\ndot}{\dot{n}}

\newcommand{\Omdot}{\dot{\Om}}

\newcommand{\clf}{{\mathcal{F}}}
\newcommand{\clg}{{\mathcal{G}}}

\newcommand{\clr}{{\mathcal{R}}}

\newcommand{\clt}{{\mathcal{T}}}

\newcommand{\da}{\partial_a}
\newcommand{\db}{\partial_b}

\newcommand{\dift}{\partial_t}

\newcommand{\si}{\bsig_1}
\newcommand{\sj}{\bsig_2}
\newcommand{\sk}{\bsig_3}

\newcommand{\isi}{I \hspace{-1pt} \bsig_1}
\newcommand{\isj}{I \hspace{-1pt} \bsig_2}
\newcommand{\isk}{I \hspace{-1pt} \bsig_3}

\newcommand{\bsi}{\bsig_1}
\newcommand{\bsj}{\bsig_2}
\newcommand{\bsk}{\bsig_3}

\newcommand{\gi}{\gamma_{1}}
\newcommand{\gj}{\gamma_{2}}
\newcommand{\gk}{\gamma_{3}}
\newcommand{\go}{\gamma_{0}}

\newcommand{\gamum}{\gamma^\mu}
\newcommand{\gamdm}{\gamma_\mu}

\newcommand{\gamdn}{\gamma_\nu}

\newcommand{\bsigph}{\bsig_\phi}

\newcommand{\Rrev}{\tilde{R}}

\newcommand{\lih}{\mathsf{h}}
\newcommand{\lbh}{\bar{\mathsf{h}}}
\newcommand{\lig}{\mathsf{g}}

\newcommand{\ho}{\lbh}
\newcommand{\hu}{\lih}

\newcommand{\liR}{\mathsf{R}}

\newcommand{\grad}{\nabla}

\newcommand{\dgrad}{\dot{\grad}}
\newcommand{\bgrad}{\mbox{\boldmath $\grad$}}

\newcommand{\deriv}[2]{\frac{\partial #1}{\partial #2}}

\newcommand{\ric}{{\cal R}}

\newcommand{\phht}{{\hat{\phi}}}

\newcommand{\dbgrad}{\dot{\bgrad}}
\newcommand{\dbn}{\mbox{\boldmath{$\ndot$}}}
\newcommand{\dtmu}{\dot{\mu}}

\newcommand{\edu}{\blde_u}
\newcommand{\euu}{\blde^u}
\newcommand{\edv}{\blde_v}
\newcommand{\euv}{\blde^v}
\newcommand{\edrho}{\blde_\rho}

\begin{document}

\vspace*{0.4cm}

\noindent
\textsf{\textbf{\large INTEGRAL EQUATIONS, KERR--SCHILD FIELDS\\ AND
GRAVITATIONAL SOURCES}} 

\vspace{0.4 cm}

\noindent
{\large Chris Doran\footnote{E-mail: c.doran@mrao.cam.ac.uk} and  Anthony
  Lasenby\footnote{E-mail: a.n.lasenby@mrao.cam.ac.uk}}

\vspace{0.4cm}
\noindent
Astrophysics Group, Cavendish Laboratory, Madingley Road, \\
Cambridge CB3 0HE, UK.

\vspace{0.4cm}

\begin{abstract}
Kerr--Schild solutions to the vacuum Einstein equations are considered
from the viewpoint of integral equations.  We show that, for a class
of Kerr--Schild fields, the stress-energy tensor can be regarded as a
total divergence in Minkowski spacetime.  If one assumes that
Minkowski coordinates cover the entire manifold (no maximal
extension), then Gauss' theorem can be used to reveal the nature of
any sources present.  For the Schwarzschild and Vaidya solutions the
fields are shown to result from a $\del$-function point source.  For
the Reissner--Nordstrom solution we find that inclusion of the
gravitational fields removes the divergent self-energy familiar from
classical electromagnetism.  For more general solutions a complex
structure is seen to arise in a natural, geometric manner with the
role of the unit imaginary fulfilled by the spacetime pseudoscalar.
The Kerr solution is analysed leading to a novel picture of its global
properties.  Gauss' theorem reveals the presence of a disk of tension
surrounded by the matter ring singularity.  Remarkably, the tension
profile over this disk has a simple classical interpretation.  It is
also shown that the matter in the ring follows a light-like path, as
one expects for the endpoint of rotating, collapsing matter.  Some
implications of these results for physically-realistic black holes are
discussed.
\end{abstract}

\section{Introduction}

Many of the important solutions to the Einstein field equations can be
expressed in Kerr--Schild form (see, for example, the discussion
in~\cite{kra-exact}).  These include all black hole solutions, and a
range of solutions representing radiation.  Here we analyse solutions
of Kerr--Schild type from the viewpoint of the gauge theory approach
to gravity~\cite{DGL98-grav,gap,DLkerr03}.  In this approach the
gravitational fields are gauge fields defined over a flat Minkowski
spacetime.  These fields ensure that all relations between physical
quantities are independent of the position and orientation of the
matter fields --- a scheme that ensures that the background spacetime
plays no dynamic role in the physics and has no measurable properties.
Kerr--Schild metrics are constructed from a null vector field in the
background Minkowski spacetime, so are particularly well-suited to
analysis via this gauge-theoretic approach.  In this paper we show
that, for all fields of Kerr--Schild type, the Einstein tensor is a
total divergence in the background Minkowski spacetime.  Various
consequences of this result are explored.  Gauss' theorem is used to
convert volume integrals of the Einstein tensor to surface integrals,
enabling us to probe the nature of the matter singularities generating
the gravitational fields.

The gauge-theory viewpoint always produces a metric that satisfies the
Einstein equations (or their generalisation to include torsion).  But
working with fields defined over a Minkowski background does place
additional restrictions on the form of the solutions.  For example,
general relativity admits two possibilities when dealing with the Kerr
solution~\cite{isr70}
\begin{enumerate}
\item
The complete Kerr manifold can be covered by a single set of Minkowski
coordinates.  This implies a discontinuity in the fields over the
entire disk region bounded by the matter singularity.
\item 
The fields are smooth everywhere away from the ring, but an observer
passing through the ring emerges in a new, asymptotically flat,
region.  This is achieved by extending the radial coordinate $r$ to
negative values, producing the maximally-extended Kerr spacetime.
\end{enumerate}
By adopting a flat-spacetime, gauge-theory formulation we restrict
ourselves to considering case~1 only.  This can be justified on the
grounds that the full, maximally-extended Kerr solution is not thought
to be a feasible endpoint for any collapse process.  Similar comments
apply to the maximally-extended Schwarzschild and Reissner--Nordstrom
solutions, neither of which are considered here.

The first applications we consider are to spherically-symmetric
fields, concentrating on the Schwarzschild, Reissner--Nordstrom and
Vaidya solutions.  In all cases the integrals provide sensible results
for the total energy contained in the fields, with the mass
contribution to the energy residing in a point-source $\del$-function.
For the Reissner--Nordstrom solution the inclusion of gravitational
fields removes the infinite electromagnetic self-energy for a point
charge familiar from classical electromagnetism~\cite{fey-lectII}.
This result is achieved without requiring any form of regularisation
procedure, and ensures that the total electromagnetic self-energy is
zero.

We next turn to more general fields following the work of Schiffer
\etal~\cite{sch73}.  These authors showed that stationary Kerr--Schild
vacuum solutions are generated by a single, complex generating
function.  This complex structure underlies the `trick' by which the
Kerr solution is obtained from the Schwarzschild solution via a
complex `coordinate transformation'~\cite{new65}.  (This is a trick
because there is no \textit{a priori} justification for expecting the
complex transformation to result in a new vacuum solution.)  The
complex structure associated with vacuum Kerr--Schild fields is shown
here to have a simple geometric origin, with the role of the unit
imaginary fulfilled by the spacetime pseudoscalar --- the same entity
that is responsible for duality transformations of the Riemann tensor.

The remainder of this paper deals with a detailed analysis of the Kerr
solution.  For this we require a careful choice of branch cut in the
complex square route in the generating function.  Once this is made,
Gauss' theorem reveals the detailed structure of the singular region,
confirming that the matter is concentrated in a ring that circulates
on a lightlike trajectory.  This is as one would expect, since the
Riemann tensor only diverges on a ring, and special relativity alone
is sufficient to predict that rotating collapsing matter will fall
inwards until its velocity becomes lightlike.  A more surprising
result is obtained from considering integrals inside the ring, which
reveal the presence of a disc of planar tension~\cite{DGL96-erice}.
This tension is isotropic over the disk and has a simple radial
dependence, rising to $\infty$ at the ring.  Remarkably, the
functional form of the tension has a simple non-gravitational
interpretation.  In non-relativistic dynamics a membrane holding
together a rotating ring of disconnected particles would be under a
constant tension.  When special-relativistic effects are included the
picture is altered by the fact that tension can act as a source of
inertia.  This introduces a radial dependence into the tension, the
functional form of which is precisely that which lies at the heart of
the Kerr solution.  These conclusions are gauge invariant and are not
artifacts of the use of the background spacetime.  This is
demonstrated by eigen-decompositions of the stress-energy and Riemann
tensors, from which we extract the gauge-invariant information.

There has been considerable debate over many years surrounding the
nature of sources for the Kerr metric.  Many physicists have attempted
to construct extended sources for which the Kerr metric could
represent the external geometry (see Krasinski for an early
review~\cite{kra78}).  More recently, a series of authors have
constructed disk sources for the Kerr metric~\cite{bic93,neu93,pic96}.
These solutions represent extended sources and do not have horizons
present.  The present work is of a different nature, dealing solely
with the structure of the singular region --- the endpoint of a
collapse process.  The first authors to consider this were Newman \&
Janis~\cite{new65} and Isreal~\cite{isr70}.  We disagree with Isreal's
result for the energy distribution over the disk, agreeing instead
with Hamity's later result~\cite{ham76}.  Our techniques enable us to
go some way beyond Hamity's description, both in revealing the
physical properties of the disk and in understanding the nature of the
singularity around the ring.  The simple physics of the disk was first
pointed out in~\cite{DGL96-erice}.

Many of the calculations here are simplified by using the language of
`spacetime algebra'~\cite{gap,hes-sta,hes-gc}.  This is crucial to
understanding the geometric nature of the complex structure at the
heart of Kerr--Schild solutions.  The algebraic structure of the
spacetime algebra is that of the Dirac $\gam$-matrices.  Using this
algebraic structure one can develop a mathematical language that is
adept at describing many aspects of relativistic physics.  This
language includes a calculus that is somewhat more powerful than any
available in alternative languages.  The gauge theory of gravity
developed in~\cite{DGL98-grav} takes on its most natural and
compelling form when expressed in the spacetime algebra.  We start
with an introduction to the spacetime algebra, giving the necessary
conventions and notations.  Further details can be found
in~\cite{DGL98-grav,gap,hes-gc,DGL95-elphys} and references contained
therein.  Reference~\cite{DGL98-grav} includes an appendix describing
how to convert between spacetime algebra and more conventional tensor
calculus.  Natural units ($G=c=\eps_0=1$) are employed throughout
this paper.

\subsection{Spacetime algebra}

The basic algebraic structure behind the spacetime algebra will be
familiar to most physicists in the guise of the algebra of the Dirac
$\gamma$-matrices.  The geometric interpretation the spacetime algebra
attaches to this algebra may be less familiar, though it is remarkably
well-suited to most problems in relativistic
physics~\cite{gap,hes-sta,DGL95-elphys}.  The spacetime algebra is
generated by four vectors $\{\gamdm\}$, $\mu=0\ldots 3$, equipped with
an associative (Clifford) product denoted by juxtaposition.  The
symmetric and antisymmetric parts of this product define the inner and
outer products, and are denoted with a dot and a wedge respectively, so
\begin{equation}
\gamdm \dt \gamdn \eqv \half (\gamdm \gamdn + \gamdn\gamdm) =
\eta_{\mu \nu} = \mbox{diag($+$\ $-$\ $-$\ $-$)} 
\end{equation}
and
\begin{equation}
\gamdm \wdg \gamdn \eqv \half (\gamdm \gamdn - \gamdn\gamdm).
\end{equation} 
The outer product of two vectors defines a bivector --- a directed
plane segment representing the plane defined by the two vectors.  A
full basis for the spacetime algebra is provided by the set
\begin{equation} 
\begin{array}{ccccc}
1 & \{\gamdm\} & \{\bsig_k, I\bsig_k\} & \{I\gamdm\} & I \\
\mbox{1 scalar} & \mbox{4 vectors} & \mbox{6 bivectors} & 
\mbox{4 trivectors} & \mbox{1 pseudoscalar}, \\
\mbox{grade 0} & \mbox{grade 1} & \mbox{grade 2} & \mbox{grade 3} &
\mbox{grade 4}
\end{array} 
\label{basis}
\end{equation}
where 
\begin{equation}
\bsig_k \eqv \gam_k \go, \quad  k=1\ldots 3
\end{equation} 
and
\begin{equation}
I\eqv\go\gi\gj\gk=\bsi\bsj\bsk.
\end{equation}
The pseudoscalar $I$ squares to $-1$, anticommutes with all odd-grade
elements and commutes with even grade elements.  Both the
$\{\bsig_k\}$ and $\{\gamdm\}$ are algebraic entities with clear
geometric significance.  They should not be thought of as matrices
acting on an internal spin space.  (The same symbols as employed in
quantum theory are used here simply because the algebraic relations
are the same.)

An arbitrary real superposition of the basis elements~\eqref{basis} is
called a `multivector' and these inherit the associative Clifford
product of the $\{\gamdm\}$ generators.  The inner and outer products
with a vector $a$ are of particular importance.  For these we write
\begin{equation}
a \dt A_r = \half(a A_r - (-1)^r A_r a), \qquad
a \wdg A_r = \half(a A_r + (-1)^r A_r a).
\end{equation} 
The outer and geometric products are associative, but the inner
product is not.  We also employ the commutator product,
\begin{equation}
A \crs B \eqv \half(AB - BA).
\end{equation} 
Vectors are usually denoted in lower case Latin, $a=a^\mu \gamdm$, or
Greek for basis frame vectors.  In the absence of brackets the inner,
outer and commutator products take precedence over geometric
products.

An inertial system is picked out by a future-pointing timelike (unit)
vector. If this is chosen to be the $\go$ direction then the
$\go$-vector determines a map between spacetime vectors
$a=a^\mu\gamdm$ and the even subalgebra of the full spacetime algebra
via
\begin{equation} 
a \go = a_0 + \ba, 
\label{1sptsplt}
\end{equation} 
where
\begin{equation} 
a_0 = a \dt \go, \hs{0.5} \mbox{and} \hs{0.5}
\ba = a \wdg \go.
\end{equation} 
The `relative vector' $\ba$ can be decomposed in the $\{\bsig_k\}$
frame and represents a spatial vector as seen by an observer in the
$\go$-frame.  Relative (or spatial) vectors in the $\go$-system are
written in bold type to record the fact that they are in fact
spacetime bivectors.  This distinguishes them from spacetime vectors,
which are left in normal type.  The $\{\bsig_k\}$ generate the (Pauli)
algebra of three-dimensional space, and we occasionally require that
the dot and wedge symbols define the three-dimensional inner and outer
products.  The convention we adopt is that, if both arguments of a dot
or wedge product are written in bold, then the product takes its
three-dimensional meaning.  For example, $\ba \wdg \bb$ is a relative
bivector, and so also a spacetime bivector, and not a spacetime
four-vector.

The vector derivative, $\grad$, is defined by
\begin{equation}
\grad \eqv \gamum \deriv{}{x^\mu}
\end{equation}
where the $\{x^\mu\}$ are a set of Cartesian coordinates and the
$\{\gamum\}$ are the reciprocal frame to the associated coordinate
frame $\{\gamdm\}$, \textit{i.e.\/} $\gamma^\mu \dt \gamma_\nu =
\delta^\mu_\nu$.  The spacetime split of the vector derivative $\grad$
goes through slightly differently, since we require that the $\bgrad$
symbol agrees with its conventional three-dimensional meaning.  This
is achieved by writing
\begin{equation} 
\go \grad = \dift + \bgrad,
\end{equation} 
so that $\bgrad=\bsig_i\partial_{x^i}$.  The $\grad$ operator has the
algebraic properties of a vector, and often acts on objects to which
it is not adjacent.  The `overdot' notation is a convenient way to
encode this:
\begin{equation} 
\dgrad A \dot{B} \eqv \gamum A \deriv{B}{x^\mu}.
\end{equation}
The $\grad$ operator acts on the object to its immediate right unless
brackets or overdots are present.  If brackets are present then
$\grad$ operates on everything in the bracket, so that, for example,
$\grad(AB)= \grad A B + \dgrad A \dot{B}$.  The same rules apply to
$\bgrad$.

One of the two gravitational gauge fields is the (position-dependent)
linear function $\hu(a)$, which maps vectors to vectors (where $a$ is
the vector argument).  Linear functions of this type have their action
extended to general multivectors via the rule
\begin{equation}
\hu(a \wdg b \cdots \wdg c) \eqv \hu(a) \wdg \hu(b) \wdg \cdots \wdg
\hu(c),
\end{equation}
which defines a grade-preserving linear operation.  The pseudoscalar
is unique up to a scale factor, and the determinant is defined by
\begin{equation}
\hu(I) = \det(\hu) I.
\end{equation}
The adjoint is denoted with an overbar, $\ho(a)$.  The function
$\hu(a)$ and its adjoint are related by~\cite{hes-gc}
\begin{align}
A_r \dt \ho(B_s) &= \ho( \hu(A_r) \dt B_s)
\qquad r \leq s  \nn \\
\hu(A_r) \dt B_s &= \hu( A_r \dt \ho(B_s))
\qquad r \geq s . 
\end{align}

A number of manipulations in linear algebra are simplified by using
the vector derivative in place of frame contractions.  For example,
the trace of $\hu(a)$ can be written as
\begin{equation}
\mbox{Tr} (\hu) = \gamum \dt \hu (\gamdm) = \da \dt \hu (a),
\end{equation}
where $\da$ is the vector derivative with respect to $a$.  The
following results are also useful:
\begin{align}
\da \, a \dt A_r &= r A_r \\
\da \, a \wdg A_r &= (n-r) A_r \\
\da A_r a &= (-1)^r (n-2r) A_r ,
\end{align}
where $A_r$ is a multivector of grade $r$ and $n$ is the dimension of
the space.

\subsection{The field equations}

The gravitational gauge fields are a linear function $\ho(a)$ mapping
vectors to vectors and a linear function $\Om(a)$ mapping vectors to
bivectors.  Both of these gauge fields have an arbitrary position
dependence.  The gauge-theoretic origin of these fields is described
in~\cite{DGL98-grav,gap}.  The gauge fields are related by the
equation
\begin{equation}
2\Om(a) = -\ho(\grad \wdg \lig(a)) + \hu^{-1}(\db) \wdg (a \dt \grad
\ho(b)), 
\end{equation}
where 
\begin{equation}
\lig(a) \eqv \ho^{-1} \hu^{-1} (a).
\end{equation}
The argument of the linear function, usually denoted by a vector $a$
or $b$, is always assumed to be independent of position.  
To recover the more conventional representation of general relativity
we introduce an arbitrary set of coordinates $x^\mu$, with $e_\mu$ the
associated coordinate frame vectors,
\begin{equation}
e_\mu \eqv \deriv{x}{x^\mu}.
\end{equation}
With $e^\mu$ denoting the reciprocal frame vectors we then define the
vectors
\begin{equation}
g_\mu = \hu^{-1}(e_\mu) , \qquad g^\mu = \ho(e^\mu).
\end{equation}
In terms of these the metric tensor is defined by
\begin{equation} 
g_{\mu\nu} = g_\mu \dt g_\nu.
\end{equation}
The $\ho(a)$ field ensures that one only ever has to make `flat-space'
contractions, which is an attractive feature of the gauge-theory
approach.

The field strength corresponding to the $\Om(a)$ gauge field is
defined by
\begin{equation} 
\liR(a\wdg b) \eqv a\dt\grad \Om(b) - b \dt \grad \Om(a) + \Om(a) \crs
\Om(b) 
\end{equation} 
and is a linear function mapping bivectors to bivectors.  From this
the covariant Riemann tensor is defined by
\begin{equation} 
\clr(a\wdg b) \eqv \liR(\hu(a \wdg b)).
\end{equation}
We often write this in the form $\clr(B)$, where $B$ is an arbitrary
(constant) bivector argument.  The tensor components of the Riemann
tensor are recovered by writing
\begin{equation}
{R^\mu}_{\nu \rho \sigma} = (g^\mu \wdg g_\nu) \dt \clr(g_\sigma \wdg
g_\rho).
\end{equation}
The Ricci and Einstein tensors are defined from the Riemann tensor in
the obvious way,
\begin{alignat}{2}
\mbox{Ricci Tensor:}& & \quad  \clr(b)&\eqv \da \dt \clr(a \wdg b) \\
\mbox{Ricci Scalar:}& & \quad \clr &\eqv \da \dt \clr(a) \\
\mbox{Einstein Tensor:}& &\quad \clg(a)&\eqv \clr(a) - \half a \clr.
\end{alignat}
Again, the tensor components of the Ricci and Einstein tensors are
easily recovered.

\subsection{Kerr--Schild fields}

We are interested in fields of the form
\begin{equation} 
\ho(a) = a + a \dt l \, l
\label{ksanz}
\end{equation}
where $l$ is a (flat-space) null vector, $l^2=0$.  This is the gauge
theory analogue of the Kerr--Schild ansatz.  The function~\eqref{ksanz}
extends to act on multivectors as
\begin{equation}
\ho(A) = \hu(A) = A + A \dt l \, l,
\end{equation}
and we see immediately that $\det(\ho)=1$.  The following results are
also useful:
\begin{gather}
\hu^{-1}(A) = \ho^{-1}(A) = A - A \dt l\, l \\
\lig(A) = A - 2 A \dt l \, l \\
\ho(l) = l.
\end{gather}
In terms of an orthonormal coordinate frame $\gamdm$ we can write
\begin{equation}
g_\mu = \gamdm - l_\mu l
\end{equation}
which confirms that the metric is given by
\begin{equation}
g_{\mu\nu} = \eta_{\mu\nu} - 2 l_\mu l_\nu,
\end{equation}
where $\eta_{\mu\nu}$ is the flat Minkowski metric tensor.

The $\Om(a)$ field defined by~\eqref{ksanz} has the simple form
\begin{align}
\Om(a) 
&= \ho\bigl( \grad \wdg (a \dt l \, l) \bigr) \nn \\
&= \grad \wdg (a \dt l \, l) - a \dt l \, v \wdg l
\label{ksom}
\end{align}
where
\begin{equation}
v \eqv l \dt \grad l.
\end{equation}
It follows from the fact that $l$ is null that 
\begin{equation}
l \dt v =0
\end{equation}
and
\begin{equation} 
\Om(l) = 0.
\end{equation}

Following the route adopted by Chandrasekhar~\cite[Section 57]{cha83},
we next form the quantity
\begin{align}
l \dt \clr(l) 
&= l \dt (\da \dt \clr(a \wdg l) ) \nn \\
&= (l \wdg \da) \dt \liR(a \wdg l) \nn \\
&= (l \wdg \da) \dt \bigl(a \dt \dgrad \dot{\Om}(l) - l \dt \grad
\Om(a)\bigr).
\end{align}
Substituting equation~\eqref{ksom} into the above we find that
\begin{align}
l \dt \clr (l)
&=  (l \wdg \da) \dt \bigl( - \dgrad((a\dt \grad l) \dt \ldot) \wdg
l  - l \dt \grad \grad \wdg (a \dt l \, l) \bigr) \nn \\
&= \da \dt l \, (a \dt \grad l) \dt v - l \dt \grad (\grad \dt l \, l
+ v) \dt l \nn \\
&= v^2 - (l \dt \grad v) \dt l \nn \\
&= 2 v^2.
\end{align}
If we are looking solely for vacuum solutions, then we can conclude
from this that $v$ must be null.  Since $v \dt l=0$, it follows that
$v$ must be parallel to $l$,
\begin{equation}
v = \phi l,
\label{leqn}
\end{equation}
where $\phi$ is a scalar field.  We will restrict attention to
solutions for which this relation does hold, even if matter is
present.  (This places a restriction on the form of matter
distributions that we can consider.)  It follows from
equation~\eqref{leqn} that $\Om(a)$ reduces to the simpler form
\begin{equation}
\Om(a) = \grad \wdg (a \dt l \, l).
\label{newOm}
\end{equation}

The Riemann tensor now splits into terms that are second-order and
fourth-order in $l$.  The fourth-order contribution is
\begin{equation}
\clr_4 (a \wdg b) = - \Omdot \bigl(((a \wdg b) \dt l \, l) \dt \dgrad
\bigr) + \Om(a) \crs \Om(b).
\label{Riem4}
\end{equation}
After some rearrangement this can be brought to the form
\begin{equation}
\clr_4 (B) = \qrt \da \dt \db \, (a \dt \grad l) lBl (b \dt \grad l) -
\qrt (a \dt \grad l) \dt (b \dt \grad l) \, \da l B l \db.
\label{clr4}
\end{equation}
Both the contraction, $\da \cdot \clr(a \wedge b)$, and the protraction,
$\da \wedge \clr(a \wedge b)$, of this contribution to the Riemann
tensor vanish.  This can be seen from the result that
\begin{equation}
\da F_1 a \wdg b F_2 = \da F_1 (ab - a \dt b) F_2 = -b F_1 F_2,
\end{equation}
which holds for any two bivectors $F_1$ and $F_2$.  The presence of
the null vector $l$ in the analogous terms in $\clr_4(B)$ ensures that
\begin{equation}
\da \clr_4 (a\wdg b) = 0,
\end{equation}
so that $\clr_4(B)$ makes no contribution to the Ricci tensor.

The only part of $\clr(B)$ that contributes to the Einstein tensor is
therefore the second-order term
\begin{equation}
\clr_2(a \wdg b) = a \dt \grad \Om(b) - b \dt \grad \Om(a).
\label{KSeq1}
\end{equation}
Contracting this and setting the result to zero we find that the
vacuum Einstein equations reduce to solving the equation
\begin{equation}
\clr(a) = \grad \dt \Om(a) - a \dt \grad \, \db \dt \Om(b) = 0.
\end{equation}

The Ricci scalar and Einstein tensor are now
straightforward to calculate:
\begin{equation}
\clr = -2 \grad \dt (\da \dt \Om(a)) 
\end{equation}
and
\begin{equation} 
\clg(a) = \grad \dt \bigl( \Om(a) - a \wdg (\db \dt \Om(b)) \bigr).
\label{Einst}
\end{equation} 
The formulae for $\Om(a)$~\eqref{newOm} and $\clg(a)$ are valid for
any Kerr--Schild type solution for which $l\dt\grad l=\phi l$.  For
such fields the Einstein tensor~\eqref{Einst} is a total divergence in
Minkowski spacetime.  In general, the field equations will be
satisfied everywhere except for some singular region over which the
fields are discontinuous.  This singular region contains the source of
the fields.  In this paper we assume that the entire solution to the
Einstein equations is described by fields defined over a single
Minkowski spacetime, so that the manifold has not been subjected to
maximal extension.  In this case we can use Gauss' theorem
straightforwardly to convert volume integrals over the source region
to surface integrals and so learn how the source matter is
distributed.  For the case of static fields, Virbhadra~\cite{vir90}
gave a formula which agrees with~\eqref{Einst} for the timelike
component $a=\go$, but the fact that the expression is a total
divergence was not exploited.

\section{Spherically-symmetric solutions}

As our first application we consider spherically-symmetric solutions.
For these it is useful to introduce a set of polar coordinates:
\begin{equation} 
\begin{alignedat}{3}
t &\eqv x \dt \go & \hs{1} \cos\! \theta &\eqv x\dt\gam^3 / r \\ 
r &\eqv \sqrt{(x \wdg \go)^2} & \hs{1} \tan\! \phi &\eqv
(x\dt\gam^2) / (x\dt\gam^1). 
\end{alignedat}
\end{equation}
We also define
\begin{equation}
e_r \eqv x \wdg \go \, \go /r, \hs{1} \bsig_r \eqv e_r \go,
\end{equation}	
and
\begin{equation} 
e_{\pm} \eqv \go \pm e_r.
\end{equation}

For spherically-symmetric solutions $l$ can be written in the form
\begin{equation}
l = \sqrt{\alp'} \, e_\pm,
\end{equation}
where $\alp'=\alp'(t,r)$.  For fields of this type it is a simple matter
to demonstrate that the fourth-order contribution to the Riemann
tensor~\eqref{clr4} vanishes.  To see this consider the case of $e_+$, for
which we obtain
\begin{align}
\clr_4(B) &= \frac{\alp'^2}{4} \bigl(- \da \dt \db \, a \dt \grad \bsig_r
(1- \bsig_r) B (1+\bsig_r) b\dt\grad \bsig_r \nn \\
&\quad + (a \dt \grad \bsig_r) \dt (b \dt \grad \bsig_r) \, \da e_+ B
e_+ \db \bigr) \nn \\ 
&=\frac{\alp'^2}{4r} \bigl( \dbgrad (1-\bsig_r)  B(1+\bsig_r)
\dot{\bsig}_r - \dbgrad (1-\bsig_r) B (1+\bsig_r) \dot{\bsig}_r \bigr) \nn
\\
&= 0,
\end{align}
with the same result holding for $e_-$.  It follows that the Riemann
tensor is given entirely by~\eqref{KSeq1}, which is also a total
divergence and so can be analysed using Gauss' theorem.  We now turn
to three applications of these results.

\subsection{The Schwarzschild solution}
\label{S-sz}

The simplest solution to the field equations is the Schwarzschild
solution, obtained from
\begin{equation}
\alp' = M/r, \hs{1} l = \sqrt{\alp'} \, (\go - e_r),
\label{swzsol}
\end{equation}
as we confirm in section~\ref{S-comp}.  The line element generated by
this solution is that of the advanced Eddington--Finkelstein form of
the Schwarzschild solution.  The Riemann tensor for the
solution~\eqref{swzsol} can be constructed using
equation~\eqref{KSeq1}, from which we find
\begin{align}
\clr(\ba) &= \ba \dt \bgrad \Om(\go) \nn \\
&= \ba \dt \bgrad \bigl( \grad \wdg (M(\go-e_r)/r) \bigr) \nn \\
&= M \ba \dt \bgrad \frac{\bx}{r^3},
\end{align}
and
\begin{align}
\clr(I\bb) &= \Omdot\bigl((I\bb) \dt \dbgrad \go \bigr) \nn \\
&= \grad \wdg \left( -\frac{M}{r} I \, \bb \wdg \bsig_r \wdg \bgrad
\, \bsig_r \go \right) \nn \\
&=  M I \, \bgrad \dt \bigl( \bb \wdg \frac{\bx}{r^3} \bigr). 
\end{align}
Away from the origin, these derivatives evaluate to
\begin{equation} 
\clr(\ba) = \frac{M}{r^3} (\ba - 3 \ba \dt \bsig_r \, \bsig_r),
\end{equation}
and
\begin{equation} 
\clr(I\bb) = \frac{I M}{r^3} (\bb - 3 \bb \dt \bsig_r \, \bsig_r),
\end{equation} 
so we can write the vacuum Riemann tensor in the form
\begin{equation}
\clr(B) = -\frac{M}{2r^3} (B + 3 \bsig_r B \bsig_r).
\label{SZ1}
\end{equation} 
Self duality of the vacuum Riemann tensor has the simple expression
$\clr(IB)=I\clr(B)$ in the spacetime algebra
formalism~\cite{DGL98-grav}).  The form of equation~\eqref{SZ1} shows
that the Riemann tensor is manifestly self-dual.  This form of the
Riemann tensor~\eqref{SZ1} for the Schwarzschild solution was first
given in~\cite{DGL-erice} and~\cite{dor-thesis}.

The form of the Riemann tensor in equation~\eqref{SZ1} is valid everywhere
away from the singularity.  To study the form of the singularity, we
return to the differential expressions for the Riemann tensor and
integrate over a sphere of radius $r_0$, centered on the origin.
Using Gauss' theorem to convert the volume integrals to surface
integrals, we obtain
\begin{equation} 
\int_{r \leq r_o} \!\!\! d^3x \, \clr(\ba) = M \int_{0}^{2 \pi} \! d\phi
\int_{0}^{\pi} \! d \theta \, \sin\! \theta \, \ba \dt \bsig_r \, \bsig_r 
= \frac{4 \pi M }{3} \ba, 
\end{equation} 
and
\begin{equation}
\int_{r \leq r_o} \!\!\! d^3x \, \clr(I\bb) = \int_{0}^{2 \pi} \! d\phi
\int_{0}^{\pi} \! d \theta \, \sin\! \theta \, I\bsig_r \dt(\bb \wdg
\bsig_r) 
= -\frac{8 \pi M }{3} I\bb.
\end{equation} 
These results combine to give
\begin{align}
\int_{r \leq r_o} \!\!\! d^3x \, \clr(B) &= \frac{4 \pi M }{3} (B \dt \go \go
-2 B\wdg \go \go) \nn \\
&= -\frac{2 \pi M }{3} (B + 3 \go B \go),
\end{align}
which contracts to yield
\begin{align}
\int d^3x \, \clr(a) &= 4 \pi M \, \go a \go  \label{SWZint1} \\
\int d^3x \, \clr &= -8 \pi M  \\
\int d^3x \, \clg(a) &= 8 \pi M a \dt \go \, \go \label{SWZint3}.
\end{align}
Since $\clr(a) = 0$ everywhere except for the origin, the integrals
\eqref{SWZint1}--\eqref{SWZint3} can be taken over any region of space
enclosing the origin.  It is clear then that the solution represents a
point source of matter, with the matter stress-energy tensor given by
\begin{equation} 
\clt(a) = M \delta(\bx) a \dt \go \go.
\label{swzste}
\end{equation} 
The same conclusion was reached in~\cite{DGL98-grav}, where the
calculations were performed in a different gauge.  This result
confirms Feynman's speculation in Lecture~15 of~\cite{fey-grav} that
``it will not be possible to demonstrate that ${G^\mu}_\nu=0$
\textit{everywhere}, but rather that ${G^\mu}_\nu=\delta(\bx)$, or
something of the kind''.  The integrals performed above are not gauge
invariant, but gauge-invariant information is extracted from them in
the form of the matter stress-energy tensor~\eqref{swzste}.
Furthermore, the integral of the Ricci scalar provides a direct
measure of the mass of the source, without the need to resort to
constructing integrals in an asymptotically flat region of spacetime.

\subsection{The Reissner--Nordstrom solution}

The Reissner--Nordstrom solution can be written in the form
\begin{equation}
\ho(a) = a + \eta a \dt e_- \, e_-
\label{RN}
\end{equation}
where $q$ the charge of the source (in natural units) and
\begin{equation}
\eta \eqv \frac{M}{r} - \frac{q^2}{8\pi r^2}.
\end{equation}
The Einstein tensor for this solution is
\begin{equation}
\clg(a) = \grad \dt \bigl( \grad (\eta a \dt e_-) \wdg e_- - \grad \dt
(\eta e_-) \, a \wdg e_- \bigr).
\end{equation}
Away from the origin we know that the mass term can be ignored, which
leaves
\begin{gather}
\clg(\go) = - \frac{q^2}{4 \pi} \grad \dt \left( \frac{\bsig_r}{r^3}
\right) = \frac{q^2}{4 \pi r^4} \, \go, \\
\clg(\gam_i) =  \frac{q^2}{8 \pi} \grad \dt \left( \grad \left(
\frac{\bsig_i \dt \bsig_r}{r^2} \right) \wdg e_- \right) = - \frac{q^2}{4 \pi
r^4} \, \bsig_r \gam_i \bsig_r.
\end{gather}
These combine to give a corresponding matter stress-energy tensor of 
\begin{equation}
\clt(a) = \frac{1}{8 \pi} \clg(a) = - \half \clf a \clf
\end{equation}
where $\clf \eqv q\bsig_r/(4\pi r^2)$.  This is the expected form for
the electromagnetic stress-energy tensor due to a point source of
charge $q$.  (See~\cite{DGL98-grav} for a detailed explanation of how to
handle electromagnetism in gauge-theory gravity.)

To study the behaviour of the fields near the origin we return to the
differential form for $\clg(a)$ and again construct integrals over a
sphere of radius $r_0$.  For this case we find that
\begin{align}
\int_{r \leq r_o} \!\!\! d^3x \, \clg(\go)  
&= \int_{r \leq r_o} \!\!\! d^3x \, \bgrad \dt \left(\frac{2M}{r^2}\,
\bsig_r - \frac{q^2}{4 \pi r^3} \, \bsig_r \right) \, \go \nn \\
&= \left(8 \pi M - \frac{q^2}{r^0} \right) \, \go, 
\end{align}
and
\begin{align}
\int_{r \leq r_o} \!\!\! d^3x \, \clg(\ba \go) 
&= \int_{r \leq r_o} \!\!\! d^3x \, \frac{q^2}{8 \pi} \go \bgrad \dt
\left(\frac{1}{r^3} \ba \wdg \bsig_r \right) \nn \\
& = \frac{q^2}{3 r_0} \, \ba \go,
\end{align}
which combine to give
\begin{equation} 
\int_{r \leq r_o} \!\!\! d^3x \, \clt(a) =  M a \dt \go \, \go +
\frac{q^2}{24 \pi r_0} (a - 4 a \dt \go \, \go).
\label{RNint}
\end{equation}
The mass term here is precisely as expected and shows again that a
point source is located at the origin. The electromagnetic
contribution is traceless, as one expects for the electromagnetic
stress-energy tensor.  Focusing attention on the $\go$-frame energy
component of the stress-energy tensor, we see that
\begin{equation}
\int_{r \leq r_o} \!\!\! d^3x \, \go \dt \clt(\go) = M - \frac{q^2}{8
\pi r_0}. 
\label{RNse}
\end{equation}
This result was also obtained by Tod~\cite{tod83}, who calculated the
quasi-local mass for the Reissner--Nordstrom solution as defined by
Penrose~\cite{pen82}.  Tod argued that this result implies that a
source for the Reissner--Nordstrom solution should have $r>q^2/(8\pi
M)$ at the surface in order to meet the dominant energy condition.
However, this misses the point that the negative contribution to the
integral comes entirely from the origin.  Everywhere off the origin
the stress-energy tensor satisfies the dominant energy condition.
Taking the integrals over the volume defined by $r_0 <r <\infty$ we
find that the electromagnetic field energy is $q^2/(8\pi r_0)$, which
agrees with the formula given by Virbhadra~\cite{vir90} and is simply
the classical result.

The electromagnetic contribution to~\eqref{RNse} is negative and finite
for all finite $r_0$, and tends to zero as the integral extends over
all space.  This is in stark contrast to the standard picture from
classical electromagnetism, where the integral of the $\go$-frame
energy $\bE^2/2$ diverges for the interior of any surface
enclosing the origin --- the classical self-energy problem discussed
by many authors (see~\cite{fey-lectII,land-fields}, for example).
Inclusion of the gravitational field has removed this divergence,
ensuring that the total electromagnetic self-energy is zero.  The
manner in which this regularisation is achieved is both simple and
instructive.  The electromagnetic energy density is rewritten as
\begin{equation}
\frac{\bE^2}{2} = \frac{q^2}{32 \pi r^4} = - \frac{q^2}{32 \pi} \bgrad
\dt \left(\frac{\bx}{r^4} \right),
\end{equation}
so that the integral over space of the electromagnetic energy density
can be converted to a surface integral, recovering the contribution
to~\eqref{RNse}.  Since the electromagnetic energy density near a point
source is very large, it is unsurprising that the inclusion of gravity
has significant consequences, and these clearly have implications for
the status of self-energies in classical field theory.  However, since
only classical fields are employed above, it is not clear whether this
result has similar implications for the divergent self-energies
encountered in QED.

\subsection{The Vaidya solution}

As a final example of the use of integral equations for
spherically-symmetric Kerr--Schild fields, we consider Vaidya's
`shining star' solution~\cite{kra-exact}.  This is generated by the 
field
\begin{equation}
\ho(a) = a + \frac{\mu(t-r)}{r} \, a \dt e_+ \, e_+ ,
\label{vd1}
\end{equation}
which is clearly similar to the Schwarzschild solution, except that
now the mass $\mu=\mu(t-r)$ is variable and the null geodesics $e_+$
are outgoing rather than incoming.  The solution~\eqref{vd1} is clearly of
Kerr--Schild type, and defining $l$ by
\begin{equation}
l =\sqrt{\mu/r}\, e_+,
\end{equation}
we find that
\begin{equation}
l \dt \grad l = \left(\frac{\mu}{r}\right)^{1/2} \, e_+ \dt \grad \, \left(
\left( \frac{\mu}{r}\right)^{1/2} e_+ \right) = - \frac{1}{2}
\left(\frac{\mu}{r^3}\right)^{1/2}  l,
\end{equation}
so that equation~\eqref{leqn} is satisfied.  The Einstein tensor
for~\eqref{vd1} is
\begin{equation}
\clg(a) = \grad \dt \left( \frac{2\mu}{r^2} a \dt e_+ \, \bsig_r \right)
\label{vdG}
\end{equation}
and away from the origin (where we can set $\grad\dt (\bsig_r/r^2)=0$)
this becomes
\begin{equation}
\clg(a) = -\frac{2\dtmu}{r^2} a \dt e_+ \, e_+,
\end{equation}
where $\dtmu= \dift \mu$.  This tensor represents a radially-symmetric
flux of outgoing massless particles.  Again, the presence of a
$\del$-function point source at the origin can be inferred from the
differential form of the Einstein tensor.  By evaluating the integral
of $\clg(a)$ over a sphere centred on the origin, and shrinking the
radius to zero, we find that
\begin{equation}
\clg(a) =  -\frac{2\dtmu}{r^2} a \dt e_+ \, e_+ + 8 \pi \mu\del(\bx) a
\dt \go \, \go. 
\end{equation}
The solution~\eqref{vd1} therefore describes a point mass at rest at
the origin which is losing mass at some arbitrary rate.  This is borne
out by the Riemann tensor,
\begin{equation}
\clr(B) = - \frac{\dtmu}{r^2} B \dt e_+ \, e_+ - \frac{\mu}{2r^3} (B +
3 \bsig_r B \bsig_r), 
\end{equation}
which exhibits a neat split into a source term describing the energy
outflow and a Weyl term due to the point mass at the origin.

The fact that the Einstein tensor is given by the divergence of a
bivector implies that
\begin{equation}
\grad \dt \clg(a) = 0.
\end{equation}
We can therefore define a conserved total energy $E$ by
\begin{align}
8\pi E &\eqv \int d^3x \, \go \dt \clg(\go) \nn \\
&= \int d^3x \, \bgrad \dt \bigl(2\mu \frac{\bsig_r}{r^2} \bigr) \nn \\
&= 8 \pi \mu(-\infty).
\end{align}
The total conserved energy is therefore determined by the mass of the
source at $t=-\infty$, before it began radiating, which is clearly a
sensible result.  A conserved energy of this form will exist for any
Kerr--Schild field of the type~\eqref{ksanz}, provided that the null vector
$l$ satisfies $l\dt\grad l = \phi l$.

\section{Stationary vacuum solutions}
\label{S-comp}

We now turn to a more general analysis, dropping the requirement of
spherical symmetry.  As we are ultimately interested in the Kerr
solution, however, we do restrict to stationary, vacuum solutions.  
For these we write $l$ in the form
\begin{equation} 
l = \sqrt{\alp'} \, n
\end{equation}
where
\begin{equation}
n = \go - \bn \go,
\end{equation} 
$\bn^2=1$, and $\alp'$ and $\bn$ are functions of the spatial position
vector $\bx = x\wdg\go$ only.  This is the most general form for a
stationary, Kerr--Schild field.  The condition that $l\dt\grad l =
\phi l$ immediately yields
\begin{equation} 
- \bn \dt \bgrad \bigl(\sqrt{\alp'}( \go - \bn \go) \bigr) = 
\phi (\go - \bn \go)
\end{equation} 
hence
\begin{equation}
\phi = - \bn \dt \bgrad \sqrt{\alp'}, 
\end{equation}
and
\begin{equation}
\bn \dt \bgrad \bn = 0.
\end{equation}
The final equation shows that the integral curves of $\bn$ are
straight lines.  These define possible incoming photon trajectories in
space.  The fact that these lines a straight in the background space
is a gauge-specific statement, and does not correspond to a
physically-observable property.

For stationary vacuum fields the equation $\clr(a)=0$ splits into the
pair of equations
\begin{equation}
\grad \dt \Om(\go) = 0 \label{3deqn1} 
\end{equation}
and
\begin{equation} 
\grad \dt \Om(\gam_i) + \bsig_i \dt \bgrad  \bigl(\bgrad \dt (\alp'
\bn) n \bigr) = 0 \label{3deqn2}.
\end{equation}
To simplify equation~\eqref{3deqn1} we need the result that 
\begin{equation} 
\Om(\go) = \grad \wdg (\alpha' n) = - \bgrad \alpha' - \bgrad \wdg
(\alpha' \bn), 
\end{equation} 
On splitting into spatial vector and bivector parts
equation~\eqref{3deqn1} reduces to
\begin{equation}
\bgrad^2 \alpha' = 0
\end{equation}
and
\begin{equation} 
\bgrad \dt \bigl(\bgrad \wdg (\alpha' \bn) \bigr) = 0.
\end{equation} 

\subsection{The hidden complex structure}

The content of the second field equation~\eqref{3deqn2} is summarised
neatly in the equation
\begin{equation}
\ba \dt \bgrad \bn + \bgrad (\ba \dt \bn) = \frac{2\alp'}{\bgrad
\dt(\alp' \bn)} (\bgrad (\ba \dt \bn)) \dt \bgrad \bn.
\end{equation}
In~\cite{sch73} the authors showed that this equation implies
that we can write
\begin{equation}
\ba \dt \bgrad \bn = \alpha \ba \wdg \bn \, \bn - I \beta
\ba \wdg \bn,
\label{3dgradan}
\end{equation} 
where $\alp$ and $\beta$ are two new real scalar functions.  We see
immediately that
\begin{equation}  
\bgrad \dt \bn = 2\alpha, \qquad \bgrad \wdg \bn = - 2 I \beta \bn, 
\label{3dgradn}
\end{equation} 
and it follows that
\begin{equation}
\bgrad \dt (\bet \bn) = 0, \hs{0.5} \implies \hs{0.5} \bn \dt \bgrad
\beta = - 2 \alp \bet. 
\end{equation}

Now, setting $\ba = \bgrad$ in equation~\eqref{3dgradan}, we obtain
\begin{equation}
\bgrad^2 \bn = \bgrad \alp - \bgrad \dt (\alp \bn) \bn - I \, \bgrad \wdg
(\beta \bn). 
\end{equation}
Similarly, writing equations~\eqref{3dgradn} in the form $\bgrad \bn =
2(\alp-I\bet \bn)$ and differentiating we obtain
\begin{equation}
\bgrad^2 \bn = 2\bgrad \alp - 2 I \bgrad(\bet \bn).
\label{del2n2}
\end{equation}
On combining these equations we find that
\begin{equation}
\bn \dt \bgrad \alp = \bet^2 - \alp^2,
\end{equation}
and we can therefore write
\begin{equation}
\bgrad \alp - I \bgrad \bet \, \bn = -(\alp^2 + \bet^2) \bn + 2 I \beta
(\alp- I\bet \bn).
\label{delgam1}
\end{equation}
This equation is simplified by employing the idempotent element $N$,
\begin{equation}
N \eqv \half n \go = \half (1-\bn),
\end{equation}
which satisfies
\begin{equation}
N^2 = N = -\bn N = -N \bn.
\end{equation}
(An idempotent is a mixed-grade multivector that squares to give
itself.  Spacetime idempotents are usually closely related to null
vectors.)  On postmultiplying by $N$, equation~\eqref{delgam1} yields
\begin{equation}
\bgrad \gamma N = \gam^2 N,
\label{3dgrgam}
\end{equation}
where
\begin{equation}
\gam \eqv \alp + I \beta.
\end{equation}
It follows that $(\bgrad \gam)^2 = \gam^4$ so that, if we define $\om$
by
\begin{equation}
\om = \frac{1}{\gam},
\end{equation}
then $\om$ must satisfy
\begin{equation}
(\bgrad\om)^2 = 1.
\end{equation}
This is the first of the pair of complex equations found
in~\cite{sch73}.  The novel feature of the derivation presented here
is that the `complex' quantity $\gam$ is of the form of a scalar $+$
pseudoscalar.  This gives a clear geometric origin to the complex
structure at the heart of the Kerr solution.  This complex structure
carries through to the form of the Riemann tensor, and hence to all of
the observable quantities associated with the solution.

Use of the idempotent $N$ simplifies many expressions and derivations.
For example, differentiating~\eqref{3dgrgam} and pre- and post-multiplying
by $N$ yields
\begin{equation} 
\bgrad^2 \gam N = 2 \gam^3 N - N \half \bigl(\gam^2 \bgrad \bn - \dbgrad
(\bgrad \gam) \, \dbn \bigr) N
\label{3ddel2a}
\end{equation} 
But we know that $\bn \dt \bgrad \gam = -\gam^2$ and $\bgrad \bn
N=2\gam N$, so we can rearrange the final term as follows:
\begin{align}
N \dbgrad (\bgrad \gam) \dbn N
&= - N \bn \dbgrad (\bgrad \gam) \dbn N \nn \\
&= N \dbgrad (\bn \bgrad \gam) \dbn N \nn \\
&= -2 N \gam^2 \bgrad \bn N - N \dbgrad \bgrad \gam
\dbn N \nn \\
&= - 2\gam^3 N.
\end{align}
This type of rearrangement is typical of the way that one can take
advantage of the properties of idempotents in the spacetime algebra.
On substituting this result into equation~\eqref{3ddel2a} we now find
that
\begin{equation}
\bgrad^2\gam \, N = 0,
\end{equation}
and hence
\begin{equation} 
\bgrad^2 \gam =0.
\end{equation} 
This is the second of the pair of complex equations found
in~\cite{sch73}.  Solving the field equations now reduces to finding a
complex harmonic function $\gam$ whose inverse $\om$ satisfies
$(\bgrad \om)^2=1$.  The above derivation reveals the geometric origin
of this complex structure, as well as demonstrating the role of the
null vector $n$ through the idempotent $N$.

To complete the solution we need to find forms for $\alp'$ and $\bn$.
For the former we note that
\begin{equation}
\frac{\bgrad \dt(\alp \bn)}{2\alp} = \frac{\alp^2 +\bet^2}{2\alp} = 
\frac{\bgrad \dt(\alp' \bn)}{2\alp'}
\end{equation}
and
\begin{equation}
\bgrad \dt \bigl( \bgrad \wdg (\alp \bn) \bigr) 
= \bgrad \dt \bigl( \bgrad \wdg (\alp' \bn) \bigr) = 0.
\end{equation}
From these it is a simple matter to show that $\alp'=M\alp$, where $M$
is some arbitrary constant.  To recover $\bn$ we use
equation~\eqref{3dgrgam} in the form
\begin{equation}
- \bgrad \om (1-\bn) = (1-\bn), \hs{1} \bgrad \om^\ast
(1+\bn) = (1+\bn) 
\end{equation}
where the $\ast$ denotes the complex conjugation operation (which can be
written as $\om^\ast=\go\om\go$).  On rearranging we obtain
\begin{gather}
(\bgrad \om + \bgrad \om^\ast) \bn = 2 + \bgrad \om - \bgrad \om^\ast \\
\implies \quad \bn = \frac{\bgrad \om + \bgrad \om^\ast - (\bgrad\om) \crs
(\bgrad \om^\ast)}{1+ \la \bgrad\om \bgrad \om^\ast\ra },
\label{defbn}
\end{gather}
where the angle brackets $\la \, \ra$ denote the projection onto the
scalar part of a multivector.  Recall here that the $\times$ symbol
represents half the commutator of the terms on either side, and not
the vector cross product.

Some further insight into the nature of the solution and the role of
the complex structure is obtained from the form of $\Om(\go)$.  From
equations~\eqref{3dgradn} to~\eqref{del2n2} it is straightforward to show that
\begin{equation}
\bgrad \wdg (\alp \bn) = I \bgrad \bet.
\end{equation}
It follows that $\Om(\go)$ is now given by
\begin{equation} 
\Om(\go) = -M ( \bgrad \alp + \bgrad \wdg (\alpha \bn)) = - M \bgrad
\gam. 
\end{equation} 
This shows how the harmonic function $\gam$ generalises the scalar
Newtonian potential.  This gives rise to many of the novel properties
of the Kerr solution.  A further result that is useful in later
calculations is that
\begin{equation}
\da \dt \Om(a) = M(\alp^2 + \bet^2) n.
\end{equation}

\subsection{The Riemann tensor}

The Riemann tensor would be expected to contain terms of order $M$ and
$M^2$, but it is not hard to see that the latter contribution
vanishes.  Using equations~\eqref{clr4} and~\eqref{3dgradan} this term
can be written in the form
\begin{equation}
\clr_4(B) = - \frac{M^2}{2} I \alp^2 \beta (1-\bn) \dbgrad B \dbn
(1+\bn). 
\end{equation}
But when $B$ is the spatial bivector $\ba$ we see that
\begin{align}
(1-\bn) \dbgrad \ba \dbn (1+\bn) 
&= (1-\bn)(2 \ba \dt \bgrad \bn - \ba \bgrad \bn)(1+\bn) \nn \\
&= (1-\bn)(2 \gam^\ast \ba \wdg \bn - 2 \gam^\ast \ba)(1+\bn) \nn \\
&= -2 \gam^\ast \ba \dt \bn (1-\bn)(1+\bn) \nn \\
&= 0,
\end{align}
and the same result holds for $B=I\bb$.  The annihilation of
orthogonal idempotents in this derivation, $(1+\bn)(1-\bn)=0$, is
merely a reexpression of the fact that $n$ is a null vector.

The only contribution to $\clr(B)$ therefore comes from $\clr_2(B)$.
For a spatial bivector this contribution can be written as
\begin{align}
\clr(\ba) = \clr(a \wdg \go) 
&= \ba \dt \bgrad \Om(\go) \nn \\
&= - M \ba \dt \bgrad \bgrad \gam \nn \\
&= - \half M \dbgrad \ba \bgrad \dot{\gam}.
\end{align}
For a vacuum solution the Riemann tensor only contains a Weyl term,
and so satisfies the self-duality property~\cite{DGL98-grav} $\clr(IB) =
I \clr(B)$.  It follows that we can write
\begin{equation}
\clr(B) = - \half M \dbgrad B \bgrad \dot{\gam},
\label{defriem}
\end{equation}
for all $B$.  This expression captures all of the terms of the Riemann
tensor in a single, highly compact expression.  Verifying that we have
a vacuum solution now reduces to the identity
\begin{equation}
\da  \dbgrad \, a \wdg b \, \bgrad \dot{\gam} = -b \bgrad^2 \gam = 0.
\end{equation}

\section{The Kerr solution}

The simplest solution to the pair of equations $\bgrad^2\gam=0$ and
$(\bgrad \om)^2=1$ is $\gam=1/r$.  This recovers the Schwarzschild
solution.  To conirm this we first see that equation~\eqref{defbn}
gives
\begin{equation} 
\bn = \bsig_r,
\end{equation}
which is the only possible vector consistent with spherical symmetry.
The null vector $n$ is given by $\go-e_r$, recovering the form of
solution analysed in section~\ref{S-sz}.  The Riemann tensor can be
found directly from equation~\eqref{defriem}, which yields
\begin{align}
\clr(B) &= \half M \bgrad (B \bx/r^3) \nn \\
&= -\frac{M}{2r^3} (B + 3 \bsig_r B \bsig_r)
\end{align}
recovering equation~\eqref{SZ1}.

Since the equations $\bgrad^2\gam=0$ and $(\bgrad \om)^2=1$ are
invariant under complex `coordinate transformations', a new solution
is obtained from the Schwarzschild solution by setting
\begin{equation}
\om = (x^2 + y^2 + (z-IL)^2)^{1/2}.
\label{Kom}
\end{equation}
This is the most general complex translation that can be applied to
the Schwarz\-schild solution~\cite{sch73} and generates the Kerr
solution.  The symbol $L$ ($L>0$) for the angular momentum is
preferred here to the more common symbol $a$ as we have already made
extensive use of $a$ as a vector variable.  As was first shown
in~\cite{sch73}, this complex transformation justifies the `trick'
first discovered by Newman \& Janis~\cite{new65}.  Precisely how the
complex square root in~\eqref{Kom} is defined is discussed further in
Section~\ref{Sing}.

From equation~\eqref{defriem}, we can immediately construct the Riemann
tensor as follows:
\begin{align}
\clr(B) &= - \frac{M}{2} \bgrad ( B \bgrad \gam) \nn \\
&= -\frac{3M}{8\om^5} \bgrad (\om^2)  B \bgrad (\om^2) +
\frac{M}{4\om^3} \bgrad ( B \bgrad(\om^2)) \nn \\
&= - \frac{M}{2 \om^3} \left( B + 3 \frac{\bx - L \, \isk}{\om} \, B \,  
\frac{\bx - L \, \isk}{\om} \right). 
\label{Riem}
\end{align}
The spacetime bivector
\begin{equation}
\bsig_\gam \eqv \frac{\bx - L \, \isk}{\om} = \bgrad \om 
\end{equation}
satisfies ${\bsig_\gam}^2=1$, so the Riemann tensor~\eqref{Riem} has the
same algebraic structure as for the Schwarzschild solution (it is
type~D).  The only difference is that the eigenvalues are now complex,
rather than real.  The Riemann tensor is only singular when $\om=0$,
which is over the ring $r=L$, $z=0$.  This is the reason why the
solution is conventionally referred to as containing a ring
singularity.

The structure of the fields away from the region enclosed by the ring
is most easily seen in an oblate spheroidal coordinate system.  Such a
system is defined by:
\begin{align} 
L \cosh\! u \, \cos\! v &= (x^2 + y^2)^{1/2} = \rho \\
L \sinh\! u \, \sin\! v &= z,
\end{align}
with $0\leq u < \infty$, $-\pi/2 \leq v \leq \pi/2$.  These relations
are summarised neatly in the single identity
\begin{equation}
L\cosh(u+Iv) = \rho + I z.
\end{equation}
The basic identities for oblate spheroidal coordinates, and their
relationship to cylindrical polar coordinates, are summarised in
Table~\ref{tabobl}

\begin{table}[t!!]
\renewcommand{\arraystretch}{1.2}
\begin{center}
\begin{tabular}{c}
\hline \hline 
\\
Cylindrical Polar Coordinates $\{\rho,\phi,z\}$ \\
\fbox{ \( \begin{array}{l}
\rho=(x^2 + y^2)^{1/2} \\
\phi = \tan^{-1}(y/x) \\
\blde_\rho = \cos\!\phi \, \bsi + \sin\!\phi \, \bsj \\
\blde_\phi = \rho \bsigph = \rho( -\sin\!\phi \, \bsi + \cos\!\phi \,
\bsj)  \\
\blde_\rho \blde_\phi \bsk = \rho I
\end{array} \) }\\
\\
Oblate Spheroidal Coordinates $\{u, \phi, v\}$ \\
\fbox{ \( \begin{array}{l}
L \cosh\! u \cos\! v = \rho \\
L \sinh\! u  \sin\! v = z \\
\edu = L(\sinh\! u \cos\! v \,\edrho + \cosh\! u \sin\! v \, \bsk) \\
\edv = L(- \cosh\! u \sin\! v \, \edrho + \sinh\! u \cos\! v \, \bsk) \\
\edu^2 = \edv^2 = L^2(\cosh^2\!u - \cos^2\!v) \\
\edu \blde_\phi \edv = \rho L^2 (\cosh^2\!u - \cos^2\!v) I
\end{array} \) } \\
\\
Further Relations \\
\fbox{ \( \begin{array}{l}
\edrho = L(\sinh\! u \cos\! v \, \euu - \cosh\! u \sin\! v \, \euv) \\
\bsk = L(\cosh\! u \sin\! v \, \euu + \sinh\! u \cos\! v \, \euv) \\
\bx = L^2 (\sinh\! u \cosh\! u \, \euu - \sin\! v \cos\! v \, \euv) \\ 
\end{array} \) }\\
\\
\hline \hline
\end{tabular}
\end{center}
\caption{Some basic relations for oblate spheroidal coordinates.}
\label{tabobl}
\end{table}

The point of adopting an oblate spheroidal coordinate system is
apparent from the form of $\om$:
\begin{align}
w &= L(\cosh^2\! u \, \cos^2\! v + \sinh^2\! u \, \sin^2\! v - 1 -2I
\sinh\! u\, \sin\! v )^{1/2} \nn \\
&= L(\sinh\! u - I \sin\! v).
\label{Kw}
\end{align}
This definition of the square root ensures that $\om\mapsto r$ at
large distances.  Equation~\eqref{defbn} yields a unit vector $\bn$ of
\begin{align}
\bn 
&= \frac{2 L \cosh\!u\, \euu - L^2(\cosh\! u \, \euu + I \cos\! v
\, \euv)  \crs (\cosh\! u \, \euu - I \cos\! v \, \euv) }{ 1+ L^2 \la
(\cosh\! u \, \euu + I \cos\! v \, \euv) (\cosh\! u \, \euu - I \cos\! v
\, \euv) \ra} \nn \\ 
&= \frac{1}{L \cosh\!u}(\edu - L\cos\! v \, \bsigph).
\end{align}
As a check,
\begin{align}
\bn \dt \bgrad \bn 
&= \left( \frac{1}{L \cosh\! u} \partial_u - \frac{1}{L \cosh^2\! u}
\partial_\phi \right) \left( \tanh\! u \cos\! v \, \edrho + \sin\! v
\, \bsk - \frac{\cos\! v}{\cosh\! u} \bsigph \right) \nn \\
&= 0
\end{align}
and
\begin{equation} 
\bn^2 = \frac{1}{\cosh^2\! u} (\cosh^2\!u - \cos^2 v + \cos^2 v) =1,
\end{equation} 
both as required.

\begin{figure}[t!]
\begin{center}
\includegraphics[height=6cm,angle=-90]{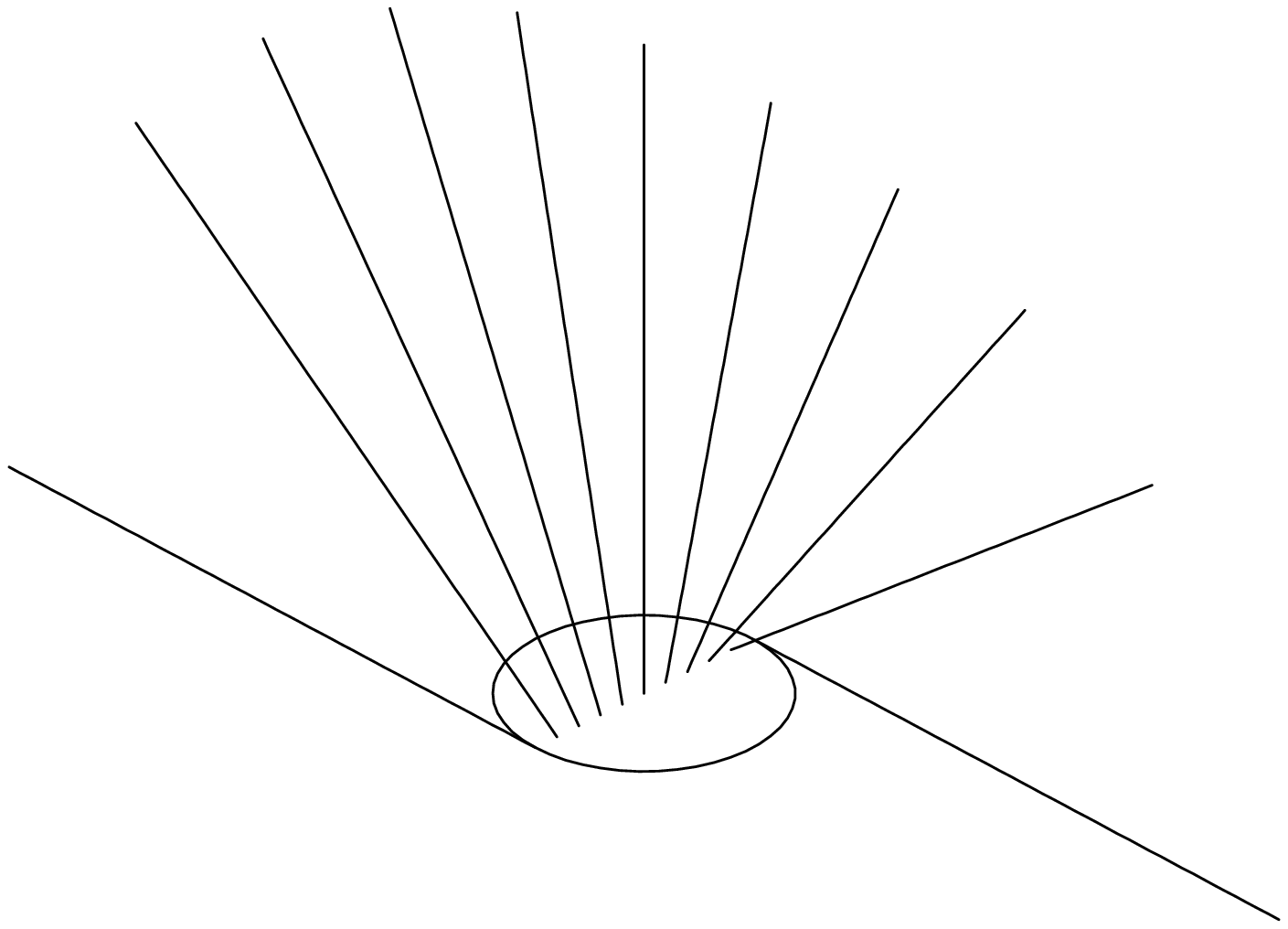}
\includegraphics[height=6cm,angle=-90]{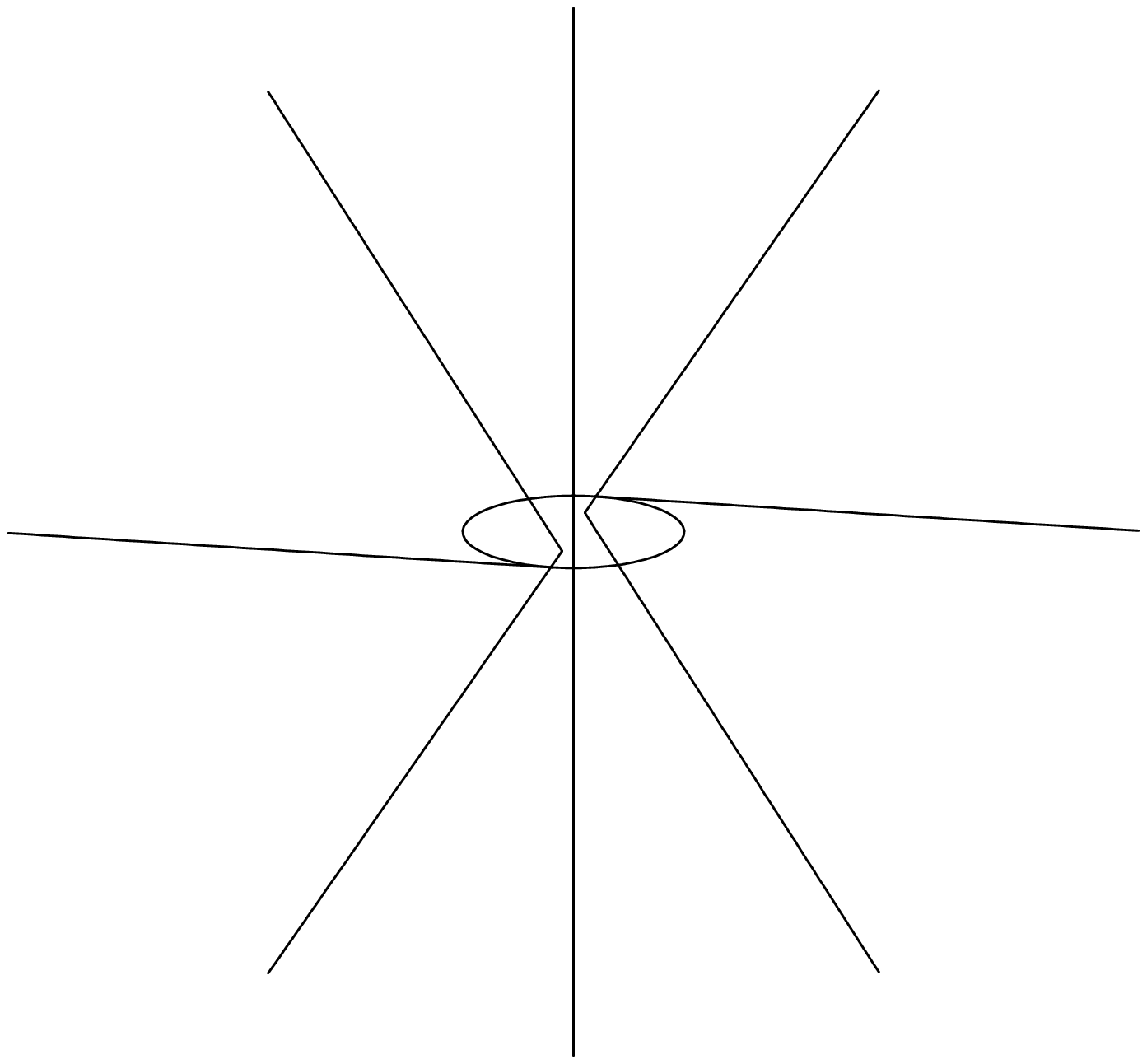}
\end{center}
\caption[dummy1]{Two views of the integral curves of $\bn$.  The
left-hand figure shows the view from above of a set of incoming
geodesics that terminate along a diameter of the central disk.  This
pattern is rotated around the $z$-axis to give the full set of
geodesics.  The right-hand figure shows incoming geodesics from above
and below the disk.  These are the mirror image of each other.  The
ring is shown for clarity --- it is not an integral curve of $\bn$.}
\label{Fig1}
\end{figure}

The vector field $\bn$ satisfies $\bn\dt\bgrad \bn=0$, so its integral
curves in flat space are straight lines.  These can be parameterised
by
\begin{equation} 
\bx(\lam) = L \cos\! v_0 \, \edrho(\phi_0) - \lam( \cos v_0 \,
\bsigph(\phi_0) - \sin\! v_0 \, \bsk),
\end{equation} 
where $L\cos\!v_0$ and $\phi_0$ are the polar coordinates for the
starting point of the integral curve over the central disk.  Plots of
these integral curves are shown in Figure~\ref{Fig1}.  As commented on
earlier, the fact that the trajectories are straight lines in the
background space is a feature of our chosen gauge.  The same picture
is not produced in alternative gauges, though the fact that the
integral curves emerge from the central disk region is gauge
invariant.

\subsection{Exterior integrals}

Oblate spheroidal coordinates are very useful for performing surface
integrals in the Kerr solution over regions entirely surrounding the
disk.  The most convenient surfaces to consider are ellipsoids of
constant $u$, for which the divergence theorem can be given in the
form 
\begin{equation} 
\int_{u' \leq u} \!\!\! d^3x \, (A \overleftrightarrow{\bgrad} B)  =
\int_{0}^{2 \pi} \!\!\! d\phi
\int_{-\pi/2}^{\pi/2} \!\!\! dv \, \rho A \edu B,
\label{divthm}
\end{equation}
where $A$ and $B$ are general multivectors.  The $\leftrightarrow$ on
$\bgrad$ indicates that the vector derivative acts both to the left
and right,
\begin{equation}
A \overleftrightarrow{\bgrad} B  = \dot{A} \dbgrad B + A (\bgrad B),
\end{equation}
and the measure $d^3x$ is taken as running over $u'$ rather than $u$.
This slightly loose notation ensures that the result of the integral
is a function of $u$.  Equation~\eqref{divthm} accounts for all of the
cases that we will encounter.

Our aim is to explore the nature of the matter singularity through the
use of integral equations.  As a first step, we look at the total
mass-energy in the source.  Taking the surface as one of constant $u$
we find that
\begin{align}
\int_{u' \leq u} \!\!\! d^3x \, \clg(\go) &= M \int_{0}^{2 \pi}
\!\!\! d\phi \int_{-\pi/2}^{\pi/2} \!\!\! dv \, \rho (\go \edu) \dt
\bigl(- \bgrad \gam + (\alp^2 + \bet^2) \bn \bigr) \nn \\
&= M \go \int_{0}^{2 \pi} \!\!\! d\phi \int_{-\pi/2}^{\pi/2} \!\!\! dv
\, \rho \left( - \deriv{\alp}{u} - \deriv{\beta}{v} \bsigph + (\alp^2 +
\bet^2) \frac{\edu^2}{L\cosh\!u} \right) \nn \\
&= 2 \pi M \go \int_{-\pi/2}^{\pi/2} \!\!\! dv \left(
\frac{\cosh^2\!u \cos\!v ( \sinh^2\!u - \sin^2\!v)}{(\sinh^2\!u +
\sin^2\! v)^2}+ \cos\!v \right) \nn \\
&= 8 \pi M \go.
\end{align}
So, as the with Schwarzschild case, the total mass-energy in the
$\go$-frame is $M$.  For the spatial part we find that
\begin{align}
\int_{u' \leq u} \!\!\! d^3x \, \clg(\ba \go) &= M \int_{0}^{2 \pi}
\!\!\! d\phi \int_{-\pi/2}^{\pi/2} \!\!\! dv \, \rho (\go \edu) \dt
\bigl( - \bgrad (\alp \ba \dt \bn) - \bgrad \wdg (\alp \ba \dt \bn \,
\bn) \nn \\ 
& \quad + (\alp^2+\bet^2)(\ba + \ba \wdg \bn) \bigr).
\end{align}
Once the angular integral is performed, the  $\go$ contribution to
this integral becomes
\begin{equation}
2 \pi M \go \int _{-\pi/2}^{\pi/2} \!\!\! dv \, \rho \bigl( - \partial_u
(\alp \sin\!v) +(\alp^2+ \bet^2)L \cosh\!u \sin\!v \bigr) = 0,
\end{equation}
which vanishes as the integrand is odd in $v$.  This is reassuring, as
we expect the integrated stress-energy tensor to be symmetric if there
is no source of torsion hidden in the singularity.  The remaining
integral to be performed transforms to
\begin{align}
\int_{u' \leq u} \!\!\! d^3x \, \clg(\ba \go) \go 
&= M \int_{0}^{2\pi} \!\!\! d\phi \int_{-\pi/2}^{\pi/2} \!\!\! dv \,
\rho I \, \edu \wdg \bgrad(\bet \ba \dt \bn) \nn \\
&= M \int_{0}^{2\pi} \!\!\! d\phi \int_{-\pi/2}^{\pi/2} \!\!\! dv \, 
I  \bigl(\rho \bsigph \, \ba \dt (\partial_v(\bet \bn)) - \edv \, \ba
\dt (\partial_\phi(\bet \bn)) \bigr) \nn \\
&= M \int_{0}^{2\pi} \!\!\! d\phi \int_{-\pi/2}^{\pi/2} \!\!\! dv \, 
I \bet \, \ba \dt \bn ( - \partial_v \rho \, \bsigph + \partial_\phi \edv)
\nn \\
&= 0.
\end{align}
In terms of the corresponding matter stress-energy tensor $\clt(a)$
the above results are summarised by 
\begin{equation}
\int d^3x \, \clt(a) = M a \dt \go \, \go,
\end{equation}
where the integral is over any region of space entirely enclosing the
central disk. 

Integrating the matter stress-energy tensor over the entire disk
region averages out any possible angular momentum contribution.  To
recover the angular momentum we look at
\begin{equation} 
x \wdg \clg(a) = t \go \wdg \clg(a) + (\bx \go) \wdg \Bigl((\go \bgrad) \dt
\bigl(\Om(a) - a \wdg (\db \dt \Om(b))\bigr) \Bigr). 
\end{equation} 
The first term on the right-hand side will give zero when integrated
over a region enclosing the disk.  To simplify the remaining term we
first write
\begin{equation} 
F(a) = \Om(a) - a \wdg (\db \dt \Om(b)).
\end{equation}
We next use the rearrangement
\begin{equation}
(\bx \go) \wdg ((\go \bgrad) \dt F(a)) = (\bx \go) \wdg ((\go
\overleftrightarrow{\bgrad}) \dt F(a)) - (F(a) \dt \go \go + 2 F(a) \wdg \go
\go),
\end{equation}
to write the volume integral as 
\begin{align}
 \int_{u' \leq u} \!\!\! d^3x \, x \wdg \clg(a)  
&= \int_{0}^{2\pi} \!\!\! d\phi \int_{-\pi/2}^{\pi/2} \!\!\! dv \,
\rho  (\bx \go) \wdg ( (\go \edu) \dt F(a)) \nn \\
& \quad  - \int_{u' \leq u} \!\!\! d^3x \, (F(a) \dt \go \go + 2 F(a)
\wdg \go \go).  
\end{align}
The final volume integral involves
\begin{equation}
\Om(a) - a \wdg (\db \dt \Om(b)) = M\grad \wdg (\alp a \dt n n) - M a
\wdg (\grad \dt (\alp n) n + \alp n \dt \grad n)
\end{equation}
which is also a total divergence and can be converted to a surface
integral.  For the $\go$ term we now find that
\begin{align}
 \int_{u' \leq u} \!\!\! d^3x \, x \wdg \clg(\go) \nn 
&=  M \int_{0}^{2\pi} \!\!\! d\phi \int_{-\pi/2}^{\pi/2} \!\!\! dv \,
\rho  \biggl(  \alp (\edu \wdg \bn \, \bn + 2 \edu \wdg \bn) \nn \\ 
& \quad +  (\bx\go) \wdg \bigl( (\go \edu) \dt (- \bgrad \gam +
(\alp^2+\bet^2) \bn) \bigr) \biggr) .
\end{align}
This simplifies down to
\begin{align}
M & \int_{0}^{2\pi} \!\!\! d\phi \int_{-\pi/2}^{\pi/2} \!\!\! dv \,
\rho \bigl(-I \, \bx \dt (\edu \wdg \bgrad \bet) + 2 \alp \edu \wdg \bn
\bigr) \nn  \\
&= - 2 \pi ML \isk \, \int_{-\pi/2}^{\pi/2} \!\!\! dv \, \left(
\frac{\cosh^2\!u \cos^3\!v (\sinh^2\!u - \sin^2\!v)}{(\sinh^2\!u +
\sin^2\! v)^2}  + \frac{2 \sinh^3\!u \cos^3\!v}{\sinh^2\!u + \sin^2\!
v} \right) \nn \\ 
&= - 8 \pi ML \, \isk.
\end{align}
For the spatial terms we obtain
\begin{multline}
\int_{u \leq u_0} \!\!\! d^3x \, x \wdg \clg(\ba\go)
= M \int_{0}^{2\pi} \!\!\! d\phi \int_{-\pi/2}^{\pi/2} \!\!\! dv \, 
\rho \biggl(  - I \, \bx \dt \bigl( \edu \wdg \bgrad (\bet
\ba \dt \bn)\bigr) \\
 + \bx \, \edu \dt \bigl( -\bgrad (\alp \ba \dt \bn) + (\alp^2 
+ \bet^2) \ba \bigr) + \alp \bigl( 2 (\edu \wdg \ba) \dt \bn \bn +
(\edu \wdg \ba) \dt \bn \bigr) \biggr), 
\end{multline}
which reduces down to
\begin{align}
& M \int_{0}^{2\pi} \!\!\! d\phi \int_{-\pi/2}^{\pi/2} \!\!\! dv \, 
\biggl(  -2L \isk \cos\!v \, \ba \dt \bn - 2 \rho \alp \,
\edu \dt \bn \, \ba \wdg \bn \nn \\
& + \rho \alp (\edu \wdg \ba) \dt \bn + \rho \bx \, \edu \dt
\bigl( - \bgrad (\alp \ba \dt \bn) + (\alp^2 + \bet^2) \ba \bigr)
\biggr) \nn \\ 
= & M \int_{0}^{2\pi} \!\!\! d\phi \int_{-\pi/2}^{\pi/2} \!\!\! dv \, 
\rho \biggl( \alp (\edu \wdg \ba) \dt \bn \nn\\
& + \bx \, \edu \dt \bigl( - \bgrad
(\alp \ba \dt \bn) + (\alp^2 + \bet^2) \ba \bigr) \biggr).
\end{align}
The final integral is best performed term by term, yielding
\begin{align}
\int d^3x \, x \wdg \clg(\gi) &= 4 \pi ML \bsj \\
\int d^3x \, x \wdg \clg(\gj) &= -4 \pi ML \bsi \\
\int d^3x \, x \wdg \clg(\gk) &= 0.
\end{align}
These results combine to give
\begin{equation}
\int d^3x \, x \wdg \clt(a) = ML \bigl( - a \dt \go \, \isk + \half (a \wdg
\go) \crs \isk \bigr),
\end{equation}
where the integral is taken over any region entirely enclosing the
central disk.  This expression clearly identifies $ML$ as the total
angular momentum in the fields, as expected from the long-range
behaviour.  The expression also has the correct algebraic form for a
symmetric stress-energy tensor.  A symmetric stress-energy tensor has
\begin{equation}
\da \wdg \clt(a) = 0.
\end{equation}
If this relation holds then we expect that
\begin{equation}
\da \wdg \bigl(- a \dt \go \, \isk + \half (a \wdg \go) \crs (\isk)
\bigr) = 0,
\end{equation}
which is easily confirmed.  We therefore expect that there are no
sources of torsion hidden in the singular region.  A discussion of the
gauge-invariance of the mass-energy and angular-momentum integrals
will be delayed until after we have a more complete understanding of
the stress-energy tensor.

\subsection{The singularity}
\label{Sing}

\begin{figure}[t!]
\begin{center}
\begin{picture}(200,160)
\put(0,160){\hbox{\includegraphics[height=7cm,angle=-90]{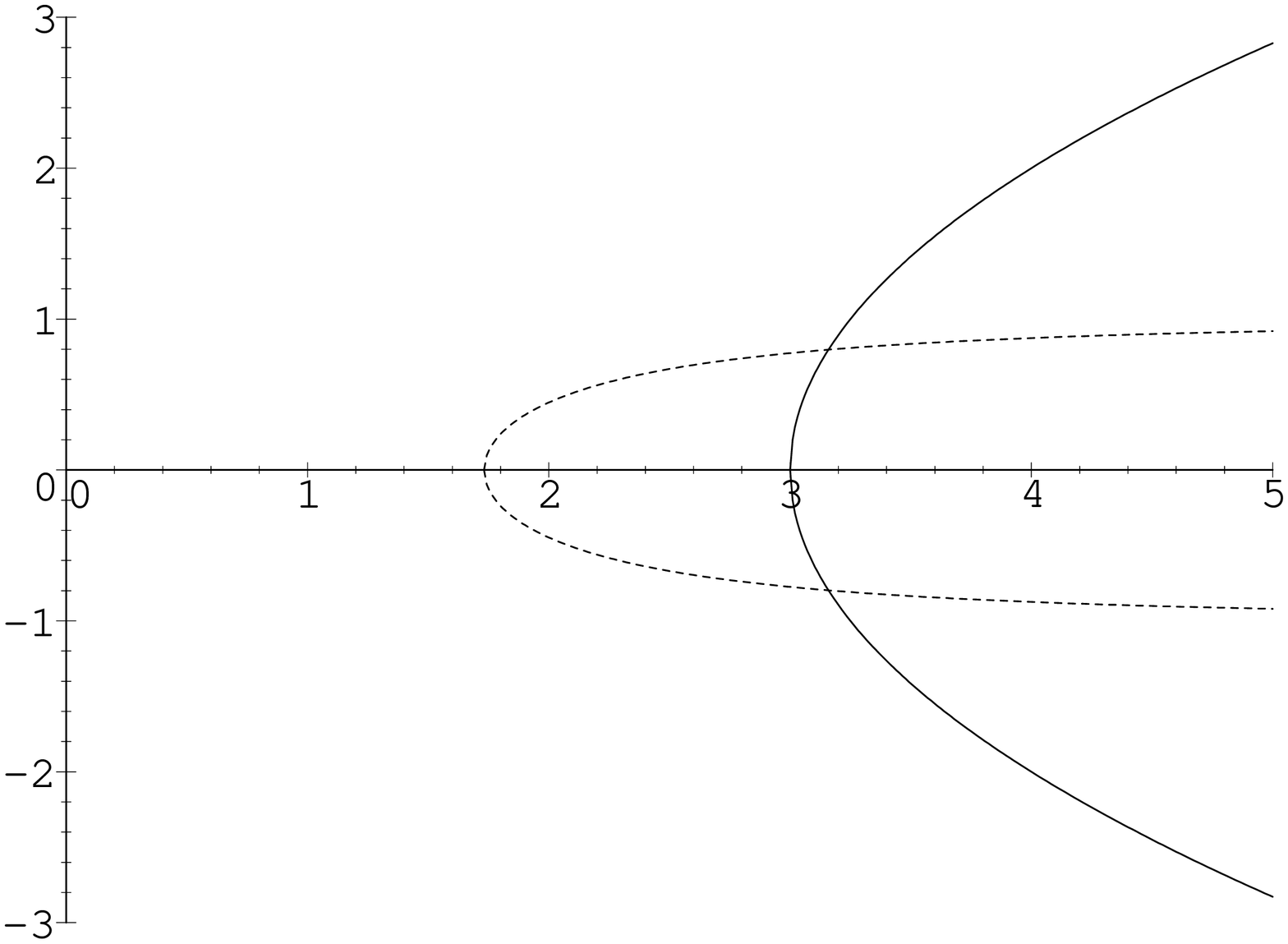}}}
\put(155,151){$\om^2(z)$}
\put(68,104){$\om(z)$}
\end{picture}
\vspace{0.5cm}
\begin{picture}(200,160)
\put(0,160){\hbox{\includegraphics[height=7cm,angle=-90]{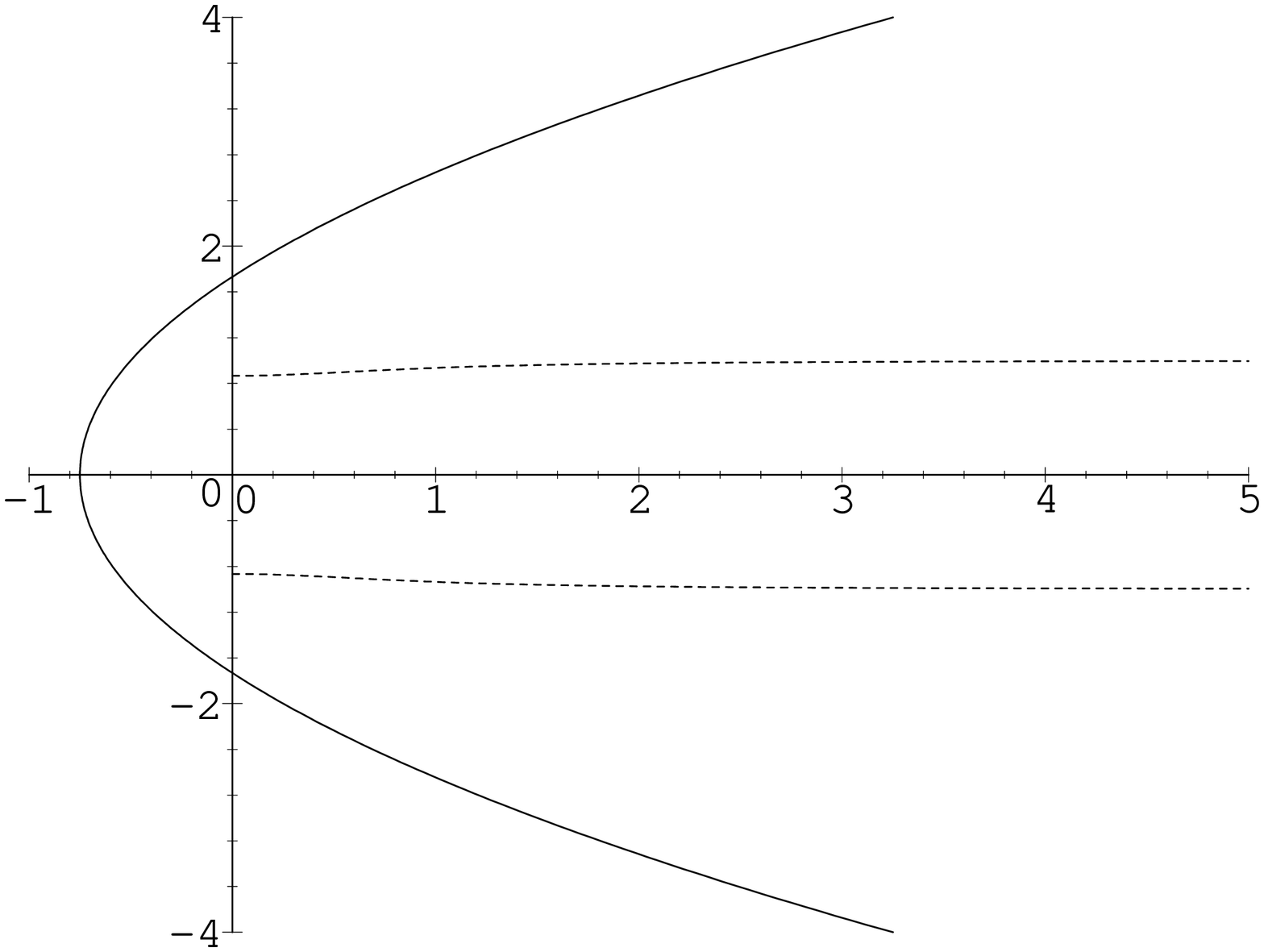}}}
\put(140,148){$\om^2(z)$}
\put(163,109){$\om(z)$}
\end{picture}
\end{center}
\caption[dummy1]{The complex function $\omega$ for fixed $\rho$ as a
function of $z$.  In both cases $L=1$.  The top figure is for $\rho =
2 >L$ and the lower figure for $\rho=0.5 < L$.  The solid lines are
for $\omega^2$ and the broken lines for the square root.  Continuity
of $\omega$ for finite $z$ requires that $\omega$ be discontiuous over
the central disk.}
\label{Figsqrt}
\end{figure}

In order to fully understand the nature of the source matter for the
Kerr solution we must look at the region $\rho < L$.  One has to be
careful with the application of oblate spheroidal coordinates in this
region, and it is safer to return to cylindrical polar coordinates for
most calculations.  Central to an understanding of this region is the
definition of the complex square root in~\eqref{Kom}. If we consider
some fixed $\rho >L$, then the complex function $\omega^2$ has a real
part $>0$ for all values of $z$ and the square root can be defined as
a smooth continuous function also with a real part $>0$ (see
Figure~\ref{Figsqrt}.$a$).  This is the definition of the square root
implicitly adopted in equation~\eqref{Kw} with the introduction of
oblate spheroidal coordinates.  If we now consider a region where
$\rho<L$ and $z$ is finite, continuity of $\om$ requires that the
square root still be defined to have a positive real part.  This means
that positive and negative $z$ now correspond to different branches of
the square root (see Figure~\ref{Figsqrt}.$b$).  As a result, $\om$ is
discontinuous across the entire disk $\rho <L$, $z=0$.  This
discontinuity is also easily seen in oblate spheroidal coordinates,
for which $z=0$, $\rho<L$ implies that $u=0$ and $\sin\!v$ is
discontinuous over the disk.  The alternative, which is not considered
here, is to extend the manifold so that passing through the disk
connects an observer to a new spacetime (a new Riemann sheet).

\subsection{The Ricci scalar}
\label{Ricci}

The simplest gauge-invariant quantity to study over the disk is the
Ricci scalar, which is given by the total divergence
\begin{equation}
\clr = -2 \grad \dt (\da \dt \Om(a)) =  - 2 M \bgrad \dt
((\alp^2+\bet^2)\bn). 
\end{equation}
We compute the integral of this over an infinite cylinder centred on
the $z$-axis of radius $\rho$, $\rho<L$.  In converting this to a
surface integral the contributions from the top and bottom of the
cylinder can be ignored, leaving
\begin{equation} 
\int_{\rho' \leq \rho} \!\!\! d^3x \, \ric =  -2 M\int_{0}^{2 \pi}
\!\!\! d\phi \int_{-\infty}^\infty \!\!\! dz \, \rho (\alp^2+\bet^2) 
\edrho \dt \bn.
\label{kerrsingint}
\end{equation} 
(As earlier, the dummy radial variable in the measure is taken as
$\rho'$, so that the result is a function of $\rho$.)  We therefore
define
\begin{equation}
W(\rho) \eqv \int_{\rho' \leq \rho} \!\!\! d^3x \, \da \dt \clt(a)/M
= \frac{1}{4 \pi} \int_{0}^{2 \pi} \!\!\! d\phi \int_{-\infty}^\infty
\!\!\! dz \, \rho \, \gam\gam^\ast \, \edrho \dt \bn.
\label{kerrsingint2}
\end{equation}

Now
\begin{equation} 
\edrho \dt \bn = \frac{\sinh\!u\, \cos\!v}{\cosh\!u}= \frac{\rho
L\sinh\!u}{L^2 \cosh^2\!u},
\end{equation} 
and we can write
\begin{equation} 
L \sinh\!u\, = {\mathbb R} \bigl((\rho^2 + (z-IL)^2)^{1/2} \bigr) =
{\mathbb R}\left(\frac{1}{\gamma} \right).  
\end{equation} 
We therefore only require an explicit expression for $L^2 \cosh^2\!u$
in terms of $\rho$ and $z$.  Such an expression is found from
\begin{align} 
\bx^2 + L^2 &= L^2(\cosh^2\!u \cos^2\!v + \sinh^2\!u \sin^2\!v
+1) \nn \\
&= L^2(\cosh^2\!u + \cos^2\!v)
\end{align} 
and
\begin{equation} 
L^2(\cosh^2\!u - \cos^2\!v)  = \bigl((\rho^2+z^2 - L^2)^2 + 4 L^2
z^2 \bigr)^{1/2},
\end{equation} 
so that
\begin{equation} 
2 L^2\cosh^2\!u\, = \rho^2+z^2 + L^2 + \bigl((\rho^2+z^2 - L^2)^2 + 4 L^2 
z^2 \bigr)^{1/2}.
\end{equation}

\begin{figure}[t!]
\begin{center}
\begin{picture}(310,300)
\put(0,300){\hbox{\epsfig{figure=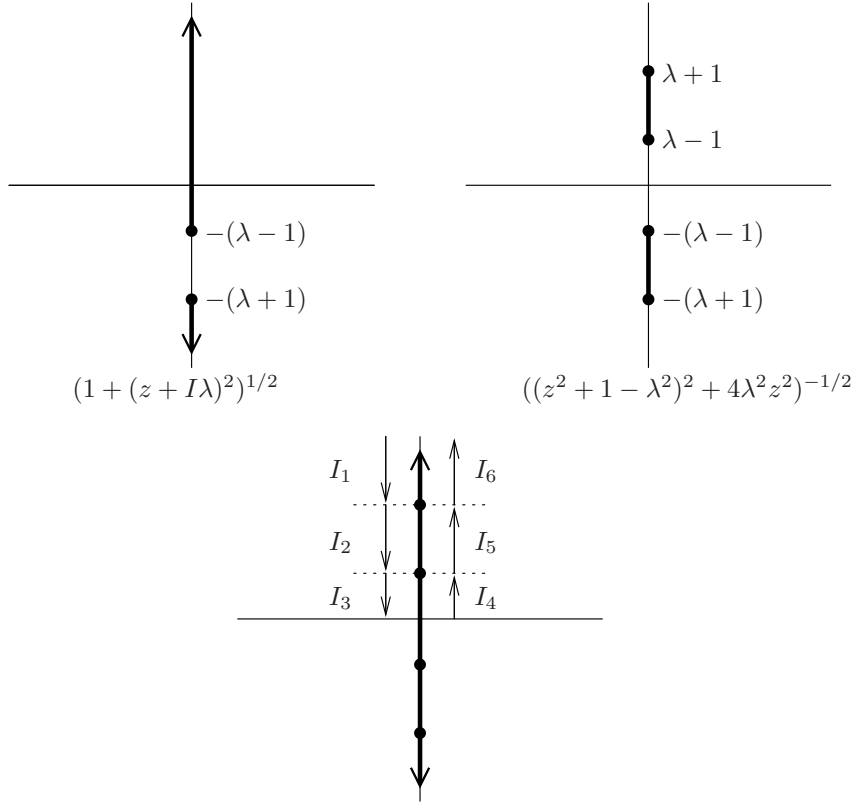,height=11cm,angle=-90}}}
\put(25,150){$(1+(z+I \lam)^2)^{1/2}$}
\put(195,150){$((z^2+1-\lam^2)^2+4\lam^2 z^2)^{-1/2}$}
\put(75,184){$-(\lam+1)$}
\put(75,209){$-(\lam-1)$}
\put(248,184){$-(\lam+1)$}
\put(248,209){$-(\lam-1)$}
\put(248,244){$\lam-1$}
\put(248,269){$\lam+1$}
\put(122,119){$I_1$}
\put(122,92){$I_2$}
\put(122,71){$I_3$}
\put(177,119){$I_6$}
\put(177,92){$I_5$}
\put(177,71){$I_4$}
\end{picture}
\end{center}
\caption[dummy1]{Branch cuts and contours for the
integral~\eqref{kerrsingint4}.  The branch cuts for complex $z$ shown in
the top two figures follow from the global definition of $\om$.  The
bottom figure shows the six integrals that have to be calculated after
the integration contour has been rotated.}
\label{Figbranch}
\end{figure}

On substituting these results into~\eqref{kerrsingint2} we obtain
\begin{equation} 
W(\rho) =\int_{-\infty}^\infty \!\!\! dz \, \rho^2 \, \frac{{\mathbb
R} \bigl(( \rho^2 + (z+IL)^2)^{-1/2} \bigr)} {\rho^2 + z^2 + L^2 +
((\rho^2+z^2 - L^2)^2 + 4 L^2 z^2)^{1/2}},
\label{kerrsingint3} 
\end{equation} 
where the integrand contains a finite jump at $z=0$, ($\rho<L$)
and no singularities.  The integral is simplified by rescaling to give
\begin{equation}
W(\rho) = \int_{-\infty}^\infty dz \, \frac{{\mathbb R}\bigl((1 +
(z+I\lambda)^2)^{-1/2}\bigr)}{z^2 +1 + \lambda^2 + ((z^2+1 - \lam^2)^2 + 4
\lambda^2 z^2)^{1/2}},
\label{kerrsingint4}
\end{equation} 
where $\lambda = L/\rho > 1$.  The branch cuts for the complex square
roots in this integral follow from the global definition of $\om$ and
are shown in Figure~\ref{Figbranch}.  The integral is performed by
splitting into the two regions $z>0$ and $z<0$ and rotating each of
the contours to lie on the positive imaginary $z$ axis.  This leaves
six integrals to compute (shown in Figure~\ref{Figbranch}) which
combine as follows:
\begin{align}
I_1 + I_6 &= -2 \int_{\lambda+1}^\infty dy \, 
\frac{((y+\lambda)^2-1)^{-1/2}}{y^2 - \lambda^2 -1 +((y^2 - \lambda^2
-1)^2-4\lambda^2)^{1/2}} \nn \\
I_2 + I_5 &= - \frac{1}{2\lambda^2} \int_{\lambda-1}^{\lambda+1} dy \,
\frac{y^2-\lambda^2-1}{((y+\lambda)^2-1)^{1/2}} \nn \\
I_3 + I_4 &= +2 \int_0^{\lambda-1} dy \,
\frac{((y+\lambda)^2-1)^{-1/2}}{\lambda^2 +1 -y^2 +((y^2 - \lambda^2
-1)^2-4\lambda^2)^{1/2}}. 
\end{align}
These combine into the simpler integrals
\begin{align}
W(\rho) &= \lim_{b \rightarrow \infty} \, - \frac{1}{2\lambda^2}
\int_0^b \!\! dy \, \frac{y^2-\lambda^2-1}{((y+\lambda)^2-1))^{1/2}} +  
\frac{1}{2\lambda^2} \int_{\lambda+1}^b \!\! dy \, ((y -
\lambda)^2-1)^{1/2} \nn \\  
& \quad - \frac{1}{2\lambda^2} \int_0^{\lambda-1} dy \, ((y -
\lambda)^2-1)^{1/2} ,
\end{align}
which are easily evaluated with cosh substitutions.  The cutoff $b$ is
introduced since the separate integrals are divergent.  On performing
the substitutions we find that 
\begin{align}
W(\rho) &= \lim_{b \rightarrow \infty} \, \frac{1}{\lambda} 
\int_{\cosh^{-1}(\lambda)}^{\cosh^{-1}(b+\lambda)} \!\! dw \, \cosh\!w\! 
- \frac{1}{2\lambda^2}
\int_{\cosh^{-1}(b-\lambda)}^{\cosh^{-1}(b+\lambda)} \!\! dw \,
\sinh^2\!w \nn \\
&= \lim_{b \rightarrow \infty} \, \frac{1}{4\lambda^2}
\bigl((b-\lambda) ((b-\lambda)^2-1)^{1/2} - (b-3\lambda)
((b+\lambda)^2-1)^{1/2} \bigr) \nn \\
& \quad - \frac{1}{\lambda} (\lambda^2-1)^{1/2} \nn \\
&= 1 - \frac{(L^2-\rho^2)^{1/2}}{L},
\label{defW}
\end{align}
so that
\begin{equation}
\int_{\rho' \leq \rho} \!\!\! d^3x \, \da \dt \clt(a) = M
\left( 1-\frac{(L^2-\rho^2)^{1/2}}{L} \right).
\label{Kintres}
\end{equation} 
Since the solution is axisymmetric, $\da \dt \clt(a)$ can only depend
on $\rho$ and $z$.  We must therefore have, for $\rho<L$,
\begin{equation}
\da \dt \clt(a) = f(\rho) \del(z),
\end{equation}
where $f(\rho)$ is found from differentiating~\eqref{Kintres}:
\begin{equation}
f(\rho) = \frac{M}{2 \pi L(L^2 - \rho^2)^{1/2}}.
\end{equation}
The function $f(\rho)$ is remarkably simple, given the convoluted
route by which it is obtained.  However, its true significance is not
seen until the remaining gauge-invariant information has been
extracted from $\clg(a)$.  This information resides in the eigenvalues
of $\clg(a)$, the calculation of which introduces further
complexities.

\subsection{The Einstein tensor}

To calculate the full form of $\clg(a)$ over the disk, we start with
the most general form that $\clg(a)$ can take consistent with the fact
that the Kerr solution is axisymmetric.  Such a form is defined by,
for $\rho<L$,
\begin{equation} 
\begin{aligned}
\clg(\go) &= \del(z) (\alp_1 \go + \bet_1 \phht + \del_1 e_\rho +
\eps_1 \gk) \\
\clg(\phht) &= \del(z) (\alp_2 \phht + \bet_2 \go + \del_2 e_\rho +
\eps_2 \gk) \\ 
\clg(e_\rho) &= \del(z) (\alp_3 e_\rho + \bet_3 \go + \del_3 \phht +
\eps_3 \gk) \\
\clg(\gk) &= \del(z) (\alp_4 \gk + \bet_4 \phht + \del_4 e_\rho +
\eps_4 \go),
\end{aligned}
\label{genEins}
\end{equation}
where each of the $\alp_i \ldots \eps_i$ are scalar functions of
$\rho$ only.  We do not assume that $\clg(a)$ is a symmetric linear
function so as to allow for the possibility that the matter contains a
hidden source of torsion.

Calculation of each of the terms in $\clg(a)$ proceeds
in the same manner as the calculation of the Ricci scalar.  The
resulting computations are long and somewhat tedious, and have been
relegated to Appendix~\ref{calcs}.  The final results are that, for
$\rho < L$,
\begin{equation} 
\begin{aligned}
\clg(\go) &= - \del(z) \frac{2M\rho}{L(L^2 - \rho^2)^{3/2}} ( \rho \go + L
\phht) \\
\clg(\phht) &= \del(z) \frac{2M}{(L^2 - \rho^2)^{3/2}} ( \rho \go + L
\phht) \\
\clg(e_\rho) &= \del(z) \frac{2M}{L(L^2 - \rho^2)^{1/2}} e_\rho \\
\clg(\gk) &= 0.
\end{aligned}
\end{equation}
These confirm that $\clg(a)$ is symmetric, so there is no hidden
torsion.  We see immediately that $e_\rho$ is an eigenvector of
$\clg(a)$, with eigenvalue $2M/(L(L^2-\rho^2)^{1/2})$, and also that
there is no momentum flow in the $\gk$ direction, which is physically
obvious.  The structure of the remaining terms is most easily seen by
introducing the boost factor $\lam$ via
\begin{equation}
\tanh\!\lam \eqv \frac{\cos\!v}{\cosh\!u}
\end{equation}
and defining the timelike velocity
\begin{equation}
v \eqv \et{\lam\bsig_\phi} \go = \cosh\!\lam \go + \sinh\!\lam
\phht. 
\end{equation}
(This second use of the symbol $v$ should not be confused with the
vector $v=l\dt\grad l$ defined earlier.)  With these definitions we
see that, over the disk,
\begin{equation}
\clg(v) = 0,
\end{equation}
and
\begin{equation}
\clg(\bsig_\phi v) =  \del(z) \frac{2M}{L(L^2 - \rho^2)^{1/2}}
\bsig_\phi v .
\end{equation}
There is therefore zero energy density in the $v$ direction, with
isotropic tension of $M/(4\pi L(L^2 - \rho^2)^{1/2})$ in the plane of the
disk.  This conclusion is gauge invariant, since it is based solely on
the eigenvalue structure of the Einstein tensor.  It is truly
remarkable that such a simple picture emerges from the complicated set
of calculations in Appendix~\ref{calcs}.  

The velocity vector $v$ defines the natural rest frame in the region
of the disk.  A second velocity is defined by the timelike Killing
vector 
\begin{equation}
g_t \eqv \hu^{-1}(\go)
\end{equation}
(see~\cite{DGL98-grav} for details of how Killing vectors are handled
within gauge-theory gravity.)  The boost required to move between
these velocities therefore provides an intrinsic definition of the
field velocity in the disk region.  In this region the function
$\ho(a)$ reduces to the identity, so $g_t$ is simply $\go$.  It
follows that the velocity is given by
\begin{equation}
\tanh \! \lam = \cos\!v = \rho/L,
\end{equation}
and the angular velocity is therefore $1/L$.  This is precisely as
expected for a rigid rotation, which fits in with the observation
of~\cite{chi92} that the Kerr solution can be viewed as the limiting
case of a rigidly-rotating matter distribution.

\subsection{An alternative gauge and the matter ring}

The form of the eigenvectors of $\clg(a)$ suggest that the more
appropriate gauge for the study of the Kerr solution is provided by
the boost 
\begin{equation}
R \eqv \et{-\lam \bsig_\phi /2}
\label{boost}
\end{equation}
so that the new solution is generated by
\begin{equation}
\ho'(a) = R(a + M \alp a \dt n \, n) \Rrev.
\label{newgauge}
\end{equation}
In this gauge the tension lies entirely in the $\isk$ eigenplane,
and the characteristic bivector of the Riemann tensor becomes
\begin{equation}
R \bsig_\gam \Rrev = \frac{\edu}{|\edu|}
\end{equation}
which is now a relative spatial vector.  In this gauge the $\go$ frame
is the rest frame defined by the Weyl and matter tensors, whereas the
Killing vectors are now swept round.

The boost~\eqref{boost} is well-defined everywhere except for the ring
singularity where the matter is located.  This is unproblematic, since
the fields are already singular there.  We know that the integral of
the Ricci scalar over the disk gives
\begin{equation}
8 \pi M \int_0^L \!\!\! d\rho \,  \frac{\rho}{L(L^2 - \rho^2)^{1/2}}
= 8 \pi M
\end{equation}
which accounts for the entire contribution to $\clr$ found for
integrals outside the disk.  It follows that the contribution to the
stress-energy tensor from the ring singularity must have a vanishing
trace, so that no contribution is made to the Ricci scalar.  It is
also clear that the entire contribution to the angular momentum must
come from the ring, since the tension is isotropic over the disk.
From these considerations it is clear that the contribution to
$\clg(a)$ from the ring singularity, in the gauge defined
by~\eqref{newgauge}, must be of the form
\begin{equation}
\clg_{\mbox{\small ring}}(a) = \frac{4 M}{L} \del(z) \, \del(\rho -L) \, a \dt
(\go + \phht) \, (\go + \phht) .
\end{equation}
This confirms that the ring current follows a lightlike trajectory ---
the natural endpoint for collapsing matter with angular momentum.  The
radius of the orbit, $L$, agrees with the minimum size allowed by
special relativity (see exercise~5.6 of~\cite{mis-grav}, for example).

\subsection{The physics of the disk}

A natural question is whether the tension field has a simple
non-gravitational explanation.  This is indeed the case.  The
equations for a special relativistic fluid are
\begin{align}
(\varepsilon + P) (v \dt \grad v) \wdg v &=  \grad P \wdg v \\
\grad \dt(\varepsilon v) &= - P \grad \dt v,
\end{align}
where $\varepsilon$ is the energy density, $P$ is the pressure and $v$
is the fluid velocity ($v^2=1$).  For the case of a ring of particles
surrounding a rigidly-rotating massless membrane under tension we see
that (ignoring the factors of $\del(z)$), $P=P(\rho)$ and
\begin{equation}
v = \cosh\!\lam \, \go + \sinh\!\lam \, \phht, \hs{1} \tanh\!\lam = \rho/L.
\end{equation}
If follows that
\begin{equation}
v \dt \grad v = \sinh^2\!\lam \, \phht \dt \grad \, \phht =
- \frac{\rho}{L^2-\rho^2} \, e_\rho,
\end{equation}
so the equation for $P$ is
\begin{equation}
\deriv{P}{\rho} -  \frac{\rho}{L^2-\rho^2} P = 0.
\end{equation}
This has the solution
\begin{equation}
P = \frac{P_0}{(L^2-\rho^2)^{1/2}},
\end{equation}
which has precisely the functional form of the tension distribution
found above.  The constant $P_0$ is found by requiring that the trace
of the stress-energy tensor returns $M$ when integrated over the disk.
This fixes the tension to $M/(4\pi L(L^2 - \rho^2)^{1/2})$, precisely
as is built into the gravitational fields.  Of course, the required
`light' membrane cannot be made from any known matter.  Indeed, the
fact that the membrane generates a tension while having zero energy
density means that it violates the weak energy condition.
Nevertheless, it is quite remarkable that such a simple physical
picture holds in a region of such extreme fields.

In writing the tension as $M/(4\pi L(L^2 - \rho^2)^{1/2})$ we are
expressing it in terms of the radial coordinate $\rho$.  This
coordinate is given physical significance by the fact that the $\ho$
function is the identity over the disk region, so $\rho$ is the proper
distance from the centre of the disk.  This gives a simple
gauge-invariant definition of $L$ as the physical radius of the disk.
Furthermore, the form of the Riemann tensor~\eqref{Riem} shows that its
complex eigenvalues are driven by $M/\om^3$, which is gauge-invariant,
and hence a physically-measurable quantity.  This affords
gauge-invariant significance to $M$ and $\om$, and hence to the
coordinates $\rho$ and $z$.  It follows that both $L$ and $M$ have
simple gauge-invariant definitions, without needing to appeal to the
asymptotic properties of the solution.

\section{Conclusions}

Many of the significant solutions to the Einstein equations can be
represented in Kerr--Schild form and gauge-theoretic approach
of~\cite{DGL98-grav} is well suited to their analysis.  For all
solutions of Kerr--Schild type where the null vector $l$ satisfies
$l\dt\grad l =\phi l$ the Einstein tensor is a total divergence in
flat spacetime.  The structure of the sources generating the fields
can therefore be elucidated by employing Gauss theorem to transform
volume integrals to surface integrals.  This approach is fully
justified within the gauge-theory formulation, since one only ever
deals with fields defined over a flat spacetime.

For the case of the Schwarzschild, Reissner--Nordstrom and Vaidya
solutions the gravitational fields are seen to result from a
$\del$-function point source of mass at the origin.  For the
Reissner--Nordstrom solution the $\del$-function point source is
surrounded by a Coulomb field.  An unexpected bonus of this approach
is that the infinite self-energy of the Coulomb field is removed by
the gravitational field.  Similar techniques can be applied to
Kinnersley's and Bonnor's work on accelerating and radiating
masses~\cite{kin69,bon94}, as will be discussed elsewhere.

Applied to more general stationary, vacuum solutions we find that the
complex structure at the heart of vacuum Kerr--Schild fields is the
same as the natural complex structure inherent in the Weyl tensor
through its self-duality symmetry.  Further algebraic insights are
obtained through the use of null vectors as idempotent elements,
simplifying many of the derivations of the vacuum equations.  Both of
these insights highlight the algebraic advantages of the spacetime
algebra approach.  A further example of this is seen clearly in
equation~\eqref{defriem}, which gives a remarkably simple and compact
expression for the Riemann tensor.

The application of Gauss' theorem to the Kerr solution reveals some
surprising features of the singularity.  The ring of matter follows a
lightlike trajectory and surrounds a disk of tension.  The tension
distribution over the disk is precisely that predicted by special
relativity.  The correct tension distribution was computed by
Hamity~\cite{ham76}, though he did not comment on its origin in terms
of classical relativistic physics.  We find no evidence of either the
negative surface energy density or the superluminal speeds claimed by
Isreal~\cite{isr70}.  Both Hamity and Isreal asserted that they used
the same results for surface layers in general relativity, but neither
gave detailed calculations, so the reason for Isreal's disagreement
with our result is hard to pin down.

Almost all trajectories in the Kerr geometry finish up on the disk,
rather than the ring.  Quite what happens when a particle encounters
the $\del$-function tension over the disk is unclear and can only
really be understood using a quantum framework to study the effect of
the disk on a wavepacket.  (A start on such an analysis in made
in~\cite{D00-kerr}.)  Assuming that all geodesics do terminate on the
disk then any non-causal features of the Kerr solution are
removed~\cite{isr70}, which is a physically attractive feature of the
picture presented here.  The fact that the tension membrane violates
the weak energy condition raises a further interesting question ---
how can it be formed from collapsing baryonic matter?  Furthermore, if
baryonic matter cannot form the membrane, then what is the endpoint of
the collapse process?  Answers to these questions will only emerge
when realistic collapse scenarios are formulated, though these are
notoriously difficult computations to perform.

The discussion in this paper implicitly rules out considering any
extensions to the manifold, such as obtained by converting the
Schwarzschild solution to Kruskal coordinates.  We therefore do not
consider distinct universes connected by Schwarzschild `throat', with
separate future and past singularities~\cite{mis-grav,haw-large}, or
the maximum analytic extension of the Reissner--Nordstrom geometry
with infinite ladder of possible `universes' connected by
wormholes~\cite{haw-large,kauf-front}.  In such scenarios the
applications of Gauss' theorem employed in this paper would not be
valid.  While infinite ladders of connected universes remain popular
with science fiction writers, there is no reason to believe they could
ever form physically in any collapse process.  The descriptions
presented here for both the Reissner--Nordstrom and Kerr solutions
have a much more plausible physical feel to them, even if the final
description of the singular region must ultimately involve quantum
gravity.  One final speculation concerns the nature of the membrane
supporting the Kerr ring singularity.  This bears a remarkable
similarity to some of the structures encountered in string theory, and
it would be of great interest to see if string theory can provide a
quantum description of such a source.

\appendix

\section{Surface integrals of the Einstein tensor}
\label{calcs}

From the form of~\eqref{genEins} there are 16 scalar functions to find
using extensions of the technique described in Section~\ref{Ricci}.
These are evaluated below.

\subsection{$\clg(\go)$}

For this term we have
\begin{equation}
\clg(\go) = \del(z) (\alp_1 \go + \bet_1 \phht + \del_1 e_\rho +
\eps_1 \gk),
\end{equation}
and $\alp_1$ and $\eps_1$ are computed directly from
\begin{equation}  
\int_{\rho' \leq \rho} \!\!\! d^3x \, \clg(\go) = 2 \pi \int_0^\rho
\!\!\! d\rho' \, \rho'\bigl(\alp_1(\rho') \go + \eps_1(\rho') \gk
\bigr). 
\end{equation}
On converting to a surface integral we obtain
\begin{align}
\int_{\rho' \leq \rho} \!\!\! d^3x \, \clg(\go) 
&= \rho \int_{0}^{2 \pi} \!\!\! d\phi \int_{-\infty}^\infty \!\!\! dz \, 
e^\rho \dt \bigl( \Om(\go) - \go \wdg (\da \dt \Om(a)) \bigr) \nn \\
&=  M \rho \int_{0}^{2 \pi} \!\!\! d\phi \int_{-\infty}^\infty \!\!\! dz \,
\left( -\partial_\rho \alp + (\alp^2 + \bet^2) \edrho \dt \bn ) \go +
\frac{1}{\rho} \partial_\phi \beta \, \gk \right) \nn \\
&= 2 \pi M \rho^2 \int_{-\infty}^\infty \!\!\! dz \, \left( {\mathbb R}
(\gam^3) + \frac{L \sinh\!u \gam \gam^\ast}{L^2 \cosh^2\!u} \right) \go,
\label{Iclggo}
\end{align}
which shows that $\eps_1 =0$.  (Here $\gamma^3$ refers to the complex
field $\gamma = 1/\om$, and not to a reciprocal frame vector.  When a
$\gamma_\mu$-vector is intended the $\gamma$ will always appear with a
subscript.)  The final term on the right-hand side of~\eqref{Iclggo}
is the integral already performed for the Ricci scalar.  For the
remaining term we need
\begin{align}
\rho^2 \int_{-\infty}^\infty \!\!\! dz \, {\mathbb R}(\gam^3) &= {\mathbb R}
\int_{-\infty}^\infty \frac{\rho^2 \,dz}{(\rho^2 + (z+IL)^2)^{3/2}} \nn \\
&= 2 - \frac{2L}{(L^2-\rho^2)^{1/2}},
\end{align}
where it is again crucial that the correct branch cuts are employed in
the evaluation of the contour integral.  Substituting this result
into~\eqref{Iclggo} we find that 
\begin{equation}  
2 \pi \int_0^\rho \!\!\! d\rho' \, \rho' \alp_1(\rho') = 4 \pi M \left( 2 -
\frac{L}{(L^2-\rho^2)^{1/2}} - \frac{(L^2 - \rho^2)^{1/2}}{L} \right),
\end{equation}
and differentiating recovers
\begin{equation}
\alp_1 = -\frac{2M\rho^2}{L (L^2 - \rho^2)^{3/2}}.
\end{equation}

For the remaining terms we first observe that
\begin{align}
\int_{\rho' \leq \rho} \!\!\! d^3x \, \rho e^\rho \dt \clg(\go) 
&= 2 \pi  \int_0^\rho \!\!\! d\rho' \, {\rho'}^2 \del_1(\rho') \nn \\ 
&=\int_{\rho' \leq \rho} \!\!\! d^3x \, (\rho e^\rho \wdg
\overleftrightarrow{\grad}) \dt \bigl(\Om(\go) - \go \wdg (\da \dt
\Om(a)) \bigr)
\nn \\
&= 0,
\label{Igorho}
\end{align}
so we must have $\del_1=0$.  To find $\bet_1$ we need to apply the
divergence theorem twice:
\begin{align}
2 \pi \int_0^\rho  \!\!\! d\rho' \, {\rho'}^2 \bet_1(\rho') &=
\int_{\rho' \leq \rho} \!\!\! d^3x \, (- \rho' \phht) \dt \clg(\go) \nn \\
&= M \rho^2 \int_{0}^{2 \pi} \!\!\! d\phi \int_{-\infty}^\infty \!\!\! dz \,
(\isk) \dt (- I \bgrad \bet) \nn \\ 
&\quad - 2 \int_{\rho' \leq \rho} \!\!\! d^3x \,
(\isk) \dt \bigl(\Om(\go) - \go \wdg (\da \dt \Om(a)) \bigr) \nn \\
&= 2 \pi M \rho^2 ( \bet(0_-) -\bet(0_+)) +2 M \int_{\rho' \leq \rho}
\!\!\! d^3x \, (\isk) \dt ( \overrightarrow{\bgrad} \dt (\alp n) n) \nn
\\
&= -\frac{4 \pi M \rho^2}{(L^2-\rho^2)^{1/2}} + 4 \pi M \rho
\int_{-\infty}^\infty \!\!\! dz  \frac{\alp \cos\!v}{\cosh\!u} \nn \\
&= -\frac{4 \pi M \rho^2}{(L^2-\rho^2)^{1/2}} + 8 \pi M L W(\rho),
\end{align}
with $W(\rho)$ as given by equation~\eqref{defW}.  This time,
differentiating yields
\begin{equation}
\bet_1 = - \frac{2M \rho}{(L^2 - \rho^2)^{3/2}}.
\end{equation}
Reassuringly, this term vanishes on the axis, as it must do for a
valid axially-symmetric solution.

\subsection{$\clg(\gk)$}

For $\clg(\gk)$ we can write
\begin{equation}
\clg(\gk) = \del(z) (\alp_4 \gk + \bet_4 \phht + \del_4 e_\rho +
\eps_4 \go). 
\end{equation}
This time we find that
\begin{align}
\int_{\rho' \leq \rho} \!\!\! d^3x \, \clg(\gk) &=
2 \pi \int_0^\rho \!\!\! d\rho' \, \rho' (\alp_4(\rho') \gk +
\eps_4(\rho') \go) \nn \\
&= -2 \pi M \rho \go \int_{-\infty}^\infty \!\!\! dz \, \partial_\rho(
\alp \bsk \dt \bn) \nn \\
& \quad + M \rho \gk \int_{0}^{2 \pi} \!\!\! d\phi \int_{-\infty}^\infty
\!\!\! dz \, (-I \bsigph) \dt (-I \bgrad (\beta \bsk \dt \bn))  \nn
\\
&= -2 \pi M \rho \go \, \partial_\rho  \int_{-\infty}^\infty \!\!\! dz \,
\alp \sin\!v \nn \\
&= 0.
\end{align}
The final term vanishes because $\alp \sin\!v$ is an odd function of
$z$.  It follows that $\alp_4= \eps_4=0$.  The same argument as at
equation~\eqref{Igorho} shows that $\del_4=0$, and for $\bet_4$ we
construct
\begin{align}
2 \pi \int_0^\rho \!\!\! d\rho' \, {\rho'}^2 \bet_4(\rho') 
&= \int_{\rho' \leq \rho} \!\!\! d^3x \, (- \rho' \phht) \dt \clg(\gk)
\nn \\
&= M\rho^2 \int_{0}^{2 \pi} \!\!\! d\phi \int_{-\infty}^\infty\!\!\!
dz \, (\isk) \dt ( - I \bgrad (\beta \bsk \dt \bn)) \nn \\
& \quad -2 \int_{\rho' \leq \rho} \!\!\! d^3x \, (\isk) \dt 
\bigl(\Om(\gk) - \gk \wdg( \da \dt \Om(a)) \bigr) \nn \\
&= -2 M \rho  \int_{0}^{2 \pi} \!\!\! d\phi
\int_{-\infty}^\infty\!\!\! dz \, (\isk) \dt (-\alp (I \bsigph) \dt n
\, n \bigr) \nn \\ 
&= 4 \pi M \rho \int_{-\infty}^\infty \!\!\! dz \, \frac{\alp \sin\!v
\cos\!v }{\cosh\!u} \nn \\
&= 0,
\end{align}
where we have again used the fact that $\sin v$ is an odd function of
$z$.  We therefore have $\clg(\gk)=0$, which is physically
reasonable.

\subsection{$\clg(e_\rho)$, $\clg(\phht)$}

For these terms we write
\begin{align}
\clg(\phht) &= \del(z) (\alp_2 \phht + \bet_2 \go + \del_2 e_\rho +
\eps_2 \gk) \\
\clg(e_\rho) &= \del(z) (\alp_3 e_\rho + \bet_3 \go + \del_3 \phht +
\eps_3 \gk).
\end{align}
The calculations are now complicated by the fact that the vector
arguments are functions of position.  To get round this problem we
must find equivalent integrals in terms of the fixed $\gi$ and $\gj$
vectors. We first form
\begin{align}
 2 \pi \int_0^\rho \!\!\! d\rho' \, {\rho'}^2 \eps_2(\rho') 
&= \int_{\rho' \leq \rho} \!\!\! d^3x \, (- \rho' \gk) \dt \clg(\phht)
\nn \\
&= M \rho^2  \int_{0}^{2 \pi} \!\!\! d\phi \int_{-\infty}^\infty\!\!\!
dz \, (-I \bsigph) \dt \bigl( \Om(\phht) - \phht \wdg (\da\dt\Om(a))\bigr) \nn \\
& \quad + \int_{\rho' \leq \rho} \!\!\! d^3x \,  (\isi) \dt \bigl(
\Om(\gi) - \gi \wdg (\da \dt \Om(a))\bigr) \nn \\
& \quad + \int_{\rho' \leq \rho} \!\!\! d^3x \, (\isj) \dt
\bigl(\Om(\gj) - \gj \wdg (\da \dt \Om(a))\bigr) \nn \\
&= - 2 \pi M \rho^2 \int_{-\infty}^\infty \!\!\!  dz \, \frac{\bet
\sinh\!u \cos\!v}{\cosh\!u} + 2 \pi M \rho \int_{-\infty}^\infty \!\!\!
dz \, \frac{\alp \sin\!v \cos\!v}{\cosh\!u} \nn \\
&= 0
\end{align}
which shows that $\eps_2=0$.  A similar calculation confirms that
$\eps_3=0$.

If $\clg(a)$ is symmetric then we expect to find $\bet_2=-\bet_1$.
This is confirmed by
\begin{align}
2 \pi \int_0^\rho \!\!\! d\rho' \, {\rho'}^2 \bet_2(\rho') 
&=  \int_{\rho' \leq \rho} \!\!\! d^3x \, \rho' \go \dt \clg(\phht)
\nn \\
&= \rho^2 \int_{0}^{2 \pi} \!\!\! d\phi \int_{-\infty}^\infty\!\!\!
dz \, \edrho \dt \bigl(\Om(\phht) - \phht \wdg (\da \dt \Om(a))\bigr) \nn \\
& \quad -  \int_{\rho' \leq \rho} \!\!\! d^3x \,  \si \dt \bigl(
\Om(\gj) - \gj \wdg (\da \dt \Om(a))\bigr) \nn \\
& \quad -  \int_{\rho' \leq \rho} \!\!\! d^3x \,   \sj \dt \bigl(
\Om(\gi) - \gi \wdg (\da \dt \Om(a))\bigr)  \nn \\
&= 2 \pi M \rho^2 \partial_\rho \int_{-\infty}^\infty\!\!\! dz \,
\frac{\alp \cos\!v}{\cosh\!u} - 2 \pi M \rho
\int_{-\infty}^\infty\!\!\! dz \, \frac{\alp \cos\!v}{\cosh\!u} \nn \\
&= 4 \pi M L \rho^2 \partial_\rho (W(\rho)/\rho) - 4 \pi M L W(\rho)
\nn \\
&= - 8 \pi M L W(\rho) + \frac{4\pi M\rho^2}{(L^2-\rho^2)^{1/2}}.
\end{align}
A similar, though slightly more involved calculation confirms that
$\bet_3=0$.

For the remaining four functions we need to consider various
combinations of integrals.  For example, 
\begin{align}
\pi \int_0^\rho \!\!\! d\rho' \, \rho' (\alp_2(\rho') + \alp_3(\rho'))
&= \int_{\rho' \leq \rho} \!\!\! d^3x \, \gam^1 \dt \clg(\gi) \nn \\
&= M \rho  \int_{0}^{2 \pi} \!\!\! d\phi \int_{-\infty}^\infty\!\!\!
dz \, \sin\!\phi (-\isk) \dt ( -I \bgrad (\beta \si \dt \bn)) \nn \\
&= \frac{\pi \rho^2 M}{L} (\bet(0_+) - \bet(0_-)) \nn \\
&= \frac{2 \pi M \rho^2}{L(L^2-\rho^2)^{1/2}},
\end{align}
from which we obtain
\begin{equation}
\alp_2 + \alp_3 = \frac{4M}{L(L^2-\rho^2)^{1/2}} + \frac{2 M
\rho^2}{L(L^2-\rho^2)^{3/2}} . 
\end{equation}
Similarly,
\begin{align}
\pi \int_0^\rho \!\!\! d\rho' \, \rho' (\del_2(\rho') - \del_3(\rho')) 
&= \int_{\rho' \leq \rho} \!\!\! d^3x \, \gam^1 \dt \clg(\gj) \nn \\
&= M \rho  \int_{0}^{2 \pi} \!\!\! d\phi \int_{-\infty}^\infty\!\!\!
dz \, \sin\!\phi \, (-\isk) \dt ( -I\bgrad (\bet \bsj \dt \bn)) \nn \\
&= 0
\end{align}
(because $\sinh\!u=0$ over the disk).  This result confirms that
$\clg(a)$ is symmetric over the disk and therefore that there are no
hidden sources of torsion.

The functions $\del_2$ and $\del_3$ are obtained from
\begin{equation}
\frac{\pi}{4} \int_0^\rho \!\!\! d\rho' \, {\rho'}^3 (\del_2(\rho') +
\del_3(\rho'))  
= \int_{\rho' \leq \rho} \!\!\! d^3x \, {\rho'}^2 \sin\!\phi
\cos\!\phi \, \gam_1 \dt \clg(\gi)
\end{equation}
which evaluates to
\begin{align}
& \quad M \rho^3  \int_{0}^{2 \pi} \!\!\! d\phi \int_{-\infty}^\infty\!\!\!
dz \, \sin^2\!\phi \cos\!\phi (\isk) \dt ( -I\bgrad (\bet \si \dt
\bn) ) \nn \\
& \quad - \int_{\rho' \leq \rho} \!\!\! d^3x \, \rho' \cos\!\phi \,
(\isk) \dt \bigl( \Om(\gi) - \gi \wdg (\da \dt \Om(a))\bigr)  \nn \\
&= - M \rho^2  \int_{0}^{2 \pi} \!\!\! d\phi \int_{-\infty}^\infty\!\!\!
dz \, \sin\!\phi \cos\!\phi \, \alp (I \, \sk \wdg \bn)^2 = 0,
\end{align}
which shows that $\del_2=\del_3=0$.  It follows that $e_\rho$ is an
eigenvector of the stress-energy tensor.  A similar trick is used to
evaluate the final term:
\begin{equation}
\frac{\pi}{4} \int_0^\rho \!\!\! d\rho' \, {\rho'}^3 (\alp_2(\rho') -
\alp_3(\rho'))  
= \int_{\rho' \leq \rho} \!\!\! d^3x \, {\rho'}^2 \sin\!\phi
\cos\!\phi \,  \gi \dt \clg(\gj) .
\end{equation}
This evaluates to
\begin{align}
&= M \rho^3 \int_{0}^{2 \pi} \!\!\! d\phi \int_{-\infty}^\infty\!\!\!
dz \, \sin^2\!\phi \cos\!\phi (\isk) \dt (-I\bgrad (\bet \bsj \dt
\bn)) \nn \\
& \quad - \int_{\rho' \leq \rho} \!\!\! d^3x \, \rho' \cos\!\phi
(\isk) \dt \bigl( \Om(\gj) - \gj \wdg(\db\ \dt \Om(b))\bigr) \nn \\
&= \frac{\pi M \rho^4}{2L(L^2-\rho^2)^{1/2}} - 2 \pi M \rho^2 W(\rho)
 + 4\pi M \int_0^\rho d\rho' \, \rho' W(\rho') , 
\end{align}
where $W(\rho)$ is as defined in equation~\eqref{defW}.  We do not need to
evaluate the final integral since we are only interested in the
derivative of the right-hand side.  This yields
\begin{equation}
\alp_2 - \alp_3 = \frac{2M\rho^2}{L(L^2-\rho^2)^{3/2}}
\end{equation}
which now gives us all of the terms in $\clg(a)$.


\begin{thebibliography}{10}

\bibitem{kra-exact}
D.~Kramer, H.~Stephani, M.~Mac{C}allum, and E.~Herlt.
\newblock {\em Exact Solutions of Einstein's Field Equations}.
\newblock Cambridge University Press, 1980.

\bibitem{DGL98-grav}
A.N. Lasenby, C.J.L. Doran, and S.F. Gull.
\newblock Gravity, gauge theories and geometric algebra.
\newblock {\em Phil. Trans. R. Soc. Lond. A}, 356:487--582, 1998.

\bibitem{gap}
C.J.L Doran and A.N. Lasenby.
\newblock {\em Geometric Algebra for Physicists}.
\newblock Cambridge University Press, 2003.

\bibitem{DLkerr03}
C.J.L. Doran and A.N. Lasenby.
\newblock New techniques for analysing axisymmetric gravitational systems {I}.
  {V}acuum fields.
\newblock {\em Class. Quant. Gravity,}, 20(6):1077--1101, 2003.

\bibitem{isr70}
W.~Isreal.
\newblock Source of the {K}err metric.
\newblock {\em Phys. Rev. D}, 2(1):641--646, 1970.

\bibitem{fey-lectII}
R.P. Feynman, R.B. Leighton, and M.~Sands.
\newblock {\em The Feynman Lectures on Physics, Volume II}.
\newblock Addison-Wesley, Reading MA, 1964.

\bibitem{sch73}
M.M. Schiffer, R.J. Adler, J.~Mark, and C.~Sheffield.
\newblock {K}err geometry as complexified {S}chwarzschild geometry.
\newblock {\em J. Math. Phys.}, 14(1):52, 1973.

\bibitem{new65}
E.T. Newman and A.I. Janis.
\newblock Note on the {K}err spinning-particle metric.
\newblock {\em J.~Math. Phys.}, 6(4):915, 1965.

\bibitem{DGL96-erice}
A.N. Lasenby, C.J.L. Doran, Y.~Dabrowski, and A.D. Challinor.
\newblock Rotating astrophysical systems and a gauge theory approach to
  gravity.
\newblock In N.~S{\'{a}}nchez and A.~Zichichi, editors, {\em Current Topics in
  Astrofundamental Physics, Erice 1996}, page 380. World Scientific, Singapore,
  1997.

\bibitem{kra78}
A.~Krasinski.
\newblock Ellipsoidal space-times, sources for the {K}err metric.
\newblock {\em Ann. Phys.}, 112:22--40, 1978.

\bibitem{bic93}
J.~Bicak and T.~Ledvinka.
\newblock Relativistic disks as sources of the {K}err metric.
\newblock {\em Phys. Rev. Lett.}, 71(11):1669--1672, 1993.

\bibitem{neu93}
G.~Neugebauer and R.~Meinel.
\newblock The {E}insteinian gravitational field of the rigidly rotating disk of
  dust.
\newblock {\em ApJ}, 414:L97, 1993.

\bibitem{pic96}
C.~Pichon and D.~Lynden-{B}ell.
\newblock New sources for the {K}err and other metrics: {R}otating relativistic
  discs with pressure support.
\newblock {\em Mon. Not. R. Astron. Soc.}, 280:1007, 1996.

\bibitem{ham76}
V.H. Hamity.
\newblock An ``interior'' of the {K}err metric.
\newblock {\em Phys. Lett. A}, 56(2):77--78, 1976.

\bibitem{hes-sta}
D.~Hestenes.
\newblock {\em Space--Time Algebra}.
\newblock {Gordon and Breach, New York}, 1966.

\bibitem{hes-gc}
D.~Hestenes and G.~Sobczyk.
\newblock {\em {C}lifford Algebra to Geometric Calculus}.
\newblock Reidel, Dordrecht, 1984.

\bibitem{DGL95-elphys}
C.J.L Doran, A.N. Lasenby, S.F. Gull, S.S. Somaroo, and A.D. Challinor.
\newblock Spacetime algebra and electron physics.
\newblock {\em Adv. Imag. \& Elect. Phys.}, 95:271, 1996.

\bibitem{cha83}
S.~Chandrasekhar.
\newblock {\em The Mathematical Theory of Black Holes}.
\newblock Oxford University Press, 1983.

\bibitem{vir90}
K.S. Virbhadra.
\newblock Energy associated with a {K}err--{N}ewman black hole.
\newblock {\em Phys. Rev. D}, 41(4):1086, 1990.

\bibitem{DGL-erice}
A.N. Lasenby, C.J.L. Doran, and S.F. Gull.
\newblock Astrophysical and cosmological consequences of a gauge theory of
  gravity.
\newblock In N.~S{\'{a}}nchez and A.~Zichichi, editors, {\em Advances in
  Astrofundamental Physics, Erice 1994}, page 359. World Scientific, Singapore,
  1995.

\bibitem{dor-thesis}
C.J.L. Doran.
\newblock {\em Geometric Algebra and its Application to Mathematical Physics}.
\newblock PhD thesis, Cambridge University, 1994.

\bibitem{fey-grav}
R.P. Feynman, F.B. Morningo, and W.G. Wagner.
\newblock {\em Feynman Lectures on Gravitation}.
\newblock Addison--Wesley, Reading MA, 1995.

\bibitem{tod83}
K.P. Tod.
\newblock Some examples of {P}enrose's quasi-local mass construction.
\newblock {\em Proc. R. Soc. Lond. A}, 388:457, 1983.

\bibitem{pen82}
R.~Penrose.
\newblock Quasi-local mass and angular momentum in general relativity.
\newblock {\em Proc. R. Soc. Lond. A}, 381:53, 1982.

\bibitem{land-fields}
L.D. Landau and E.M. Lifshitz.
\newblock {\em The Classical Theory of Fields. (Fourth Edition)}.
\newblock Pergamon Press, 1975.

\bibitem{chi92}
F.J. Chinea and L.M. Gonzalez-Romero.
\newblock A differential form approach for rotating perfect fluids in general
  relativity.
\newblock {\em Class. Quantum Grav.}, 9:1271, 1992.

\bibitem{mis-grav}
C.W. Misner, K.S. Thorne, and J.A. Wheeler.
\newblock {\em Gravitation}.
\newblock W.H. Freeman and Company, San Francisco, 1973.

\bibitem{kin69}
W.~Kinnersley.
\newblock Field of an arbitrarily accelerating point mass.
\newblock {\em Phys. Rev.}, 186(5):1335, 1969.

\bibitem{bon94}
W.~B. Bonner.
\newblock The photon rocket.
\newblock {\em Class. Quantum Grav.}, 11:2007, 1994.

\bibitem{D00-kerr}
C.J.L. Doran.
\newblock New form of the {K}err solution.
\newblock {\em Phys. Rev. D}, 61(6):067503, 2000.

\bibitem{haw-large}
S.W. Hawking and G.F.R. Ellis.
\newblock {\em The Large Scale Structure of Space-Time}.
\newblock Cambridge University Press, 1973.

\bibitem{kauf-front}
W.J. Kaufmann.
\newblock {\em The Cosmic Frontiers of General Relativity}.
\newblock Penguin Books, 1979.

\end{thebibliography}
\end{document}